\documentclass[10pt,a4]{article}
\usepackage{mathrsfs,amsmath,amsthm,latexsym,amssymb,amsfonts,epsfig,dsfont,txfonts,undertilde,cancel}
\newcommand{\makeSymbol}[1]{\mathord{\vcenter{\hbox{#1}}}}
\usepackage{hyperref}
\hypersetup{bookmarks=true, bookmarksnumbered=true,
bookmarksopen=true, colorlinks, linkcolor=red, citecolor=magenta,
plainpages=false, pdfstartview=FitH}
\numberwithin{equation}{section}
\usepackage{appendix}
\allowdisplaybreaks
\begin{document}

\title{Graphical method in loop quantum gravity: II. The Hamiltonian constraint and inverse volume operators}

\author{Jinsong Yang$^{1,2}$\thanks{yangksong@gmail.com}, Yongge Ma$^3$\thanks{Corresponding author, mayg@bnu.edu.cn} \vspace{0.5em}\\
$^1${\small Department of Physics, Guizhou University, Guiyang 550025, China} \\
$^2${\small Institute of Physics, Academia Sinica, Taiwan}\\
$^3${\small Department of Physics, Beijing Normal University, Beijing 100875, China}}
\date{}

\maketitle

\begin{abstract}
This is the second paper in the series to introduce a graphical method to loop quantum gravity. We employ the graphical method as a powerful tool to calculate the actions of the Euclidean Hamiltonian constraint operator and the so-called inverse volume operator on spin network states with trivalent vertices. Both of the operators involve the co-triad operator which contains holonomies by construction. The non-ambiguous, concise and visual characters of our graphical method ensure the rigour for our calculations. Our results indicate some corrections to the existing results in literatures.
\end{abstract}

{\hspace{1em}PACS numbers: 04.60.Pp, 04.60.Ds}

\section{Introduction}

It is well known that quantum dynamics is a central issue in loop quantum gravity (LQG) (see \cite{Ashtekar:2004eh,Han:2005km} for review arcticles, and \cite{Rovelli:2004tv,Thiemann:2007bk} for books). There are two main approaches to the quantum dynamics, based on the canonical and covariant quantization programs respectively. In canonical quantization, the quantum dynamics is determined by some quantum Hamiltonian constraint operator. In the covariant program the quantum dynamics is to define a reasonable transition amplitude. One expects that the quantum dynamics from the two different approaches can make the same physical predictions. Such an expectation has been achived at least in 3-dimensional LQG to certain sense \cite{Noui:2004iy}. Although some progress has been made for 4-dimensional case in checking the consistency between the two approaches \cite{Alesci:2010gb,Alesci:2011ia,Thiemann:2013lka}, the issue  is not yet understood up to now. To understand the relation between the canonical and covariant quantum dynamics, we not only need a suitable definition of the Hamiltonian constraint operator, but also have to calculate its matrix elements on given quantum states. In the light of the seminal work by Thiemann \cite{Thiemann:1996aw,Thiemann:1997rt}, some mathematically well-defined Hamiltonian constraint operators have been constructed in LQG. For instance, a Hamiltonian constraint operator alternative to that in \cite{Thiemann:1996aw} was proposed in \cite{Borissov:1997ji} and its matrix elements acting on a gauge-invariant trivalent vertex was derived by the graphical Penrose binor calculus. Later on, the matrix elements was re-derived in \cite{Alesci:2010gb}, and then the formula in \cite{Alesci:2010gb} was corrected by sign factors in \cite{Alesci:2011ia} using graphical method. Matter coupling is also an important issue in LQG. In the case of gravity coupled to a scalar field, the whole Hamiltonian constraint operator was constructed \cite{Thiemann:1997rt,Han:2006iqa}. The matter part of the whole Hamiltonian constraint operator usually contains the ``inverse volume operator'', which is defined by the co-triad operators. In the symmetric model of loop quantum cosmology (LQC) \cite{Ashtekar:2003hd}, the analogue of the inverse volume operator is bounded above. This fact is sometimes thought as a reason for the singularity resolution in LQC. In particular, it is shown in \cite{Gupt:2011jh} that in spatially curved anisotropic models inverse volume effects may becomes important to bind expansion and shear scalars. However, it is shown in \cite{Brunnemann:2005ip} that the inverse volume operator with certain ordering in full LQG is unbounded on the zero volume eigenstates (at a gauge-invariant trivalent vertex). This throws doubt on whether one can generalize the conclusions of LQC to LQG. To definitely understand the inverse volume operator in LQG and its relation to the analogues in certain symmetric models, it is necessary to calculate in details its action on the quantum states in LQG. There is no doubt that a simple and practical calculation method is desirable to further understand both the inverse volume and the Hamiltonian constraint operators.

Note that the matrix elements of the Hamiltonian constraint operator were calculated in \cite{Borissov:1997ji} and \cite{Alesci:2010gb} by the graphical Penrose binor method and the modified Brink's graphical method respectively, while the inverse volume operator was calculated in \cite{Brunnemann:2005ip} by the algebraic method. Although the graphical Penrose binor calculus looks simpler and more intuitive than the algebraic calculation, the rules of transforming graphs were not shown in \cite{Borissov:1997ji}. So it is not obvious whether the rules correspond uniquely to the algebraic manipulations of the formula. On the other hand, a detailed derivation of the matrix elements of the Hamiltonian constraint operator was not presented in \cite{Alesci:2010gb}.

The graphical method developed by Brink in \cite{brink1968angular} was extended and introduced to LQG (see e.g. \cite{graph-I}) for a simple and non-ambiguous calculation method. This method consists of two ingredients, graphical representation and graphical calculation. The algebraic formula can be represented by the corresponding graphical formula in an unique and unambiguous way. Then one can do calculations following the simple rules of transforming graphs, corresponding uniquely to the algebraic manipulation of the formula. The graphical method displayed a powerful efficiency in the derivation of the matrix element of the volume operator which involves only the flux operator in \cite{graph-I}.  In this paper, we will consider the actions of the Euclidean gravitational Hamiltonian constraint operator and the inverse volume operator on spin network states. Both operators depend also on the honolomies in addition to fluxes.  Our aim is in two folds. One is to show that our graphical method is suitable to calculating the actions of different kinds of operators on spin network states. The other is to cross-check the results obtained in literatures, on which some important applications are based. Note that, In order to obtain the on-shell anomaly-free quantum constraint algebra, one has to employ degenerate triangulation at the co-planar vertices of spin networks in the regularization procedure of Thiemann's Hamiltonian. This problem has been overcome by a new proposed Hamiltonian constrain operator in \cite{Yang:2015zda}. 

This paper is divided into four sections. In section \ref{sec-II}, we will give a brief review of the construction of Thiemann's Euclidean Hamiltonian constraint operator, and then calculate its action on gauge invariant trivalent spin network states by our graphical method. In section \ref{sec-III}, we compute the action of the inverse volume operator appeared in the Hamiltonian constraint for gravity coupled to a scalar field. The results are summarized in section \ref{sec-IV}.

\section{The Hamiltonian constraint operator}\label{sec-II}
To quantize a classical function, we need to first regularize it into a formula represented by the fundamental variables, then to replace the variables by their quantum operators and thus obtain a regularized quantum operator. Finally, we remove regulator by a suitable limit and obtain the corresponding quantum operator. In this section, we first recall the construction of Thiemann's Hamiltonian constraint operator, and then derive the action of the Euclidean Hamiltonian constraint operator on a spin network function over trivalent vertices.

\subsection{Quantization of the Hamiltonian constraint}

The classical Hamiltonian constraint of pure gravity in the connection formulation of general relativity is given by
\begin{align}
H(N)&=\frac{1}{2\kappa}\int {\rm d}^3x\,N\frac{\tilde{E}^a_i\tilde{E}^b_j}{\sqrt{\det(q)}}\left[\epsilon_{ijk}F^k_{ab}-2(1+\beta^2)K^i_{[a}K^j_{b]}\right]=:H^E(N)-2(1+\beta^2)T(N)\,,
\end{align}
where $\kappa=8\pi G$, $\beta$ is the Barbero-Immirzi parameter, $F^k_{ab}$ is the curvature of $SU(2)$ connection $A^i_c$, the sensitized triad $\tilde{E}^b_j$ is the conjugate momentum of $A^i_c$, $K^i_ae_{bi}$ is the extrinsic curvature of a spatial hypersurface $\Sigma$ in a spacetime with cotriad $e_{bi}$, and $\det(q)$ denotes the determinant of the 3-metric $q_{ab}$. The function $H^E(N)$ is called the Euclidean Hamiltonian constraint. In what follows, we focus on the regularization of $H^E(N)$. Let us triangulate $\Sigma$ into tetrahedra $\Delta$ so that the above integral becomes a sum of integrals over $\Delta$, i.e., $\int_\Sigma=\sum_\Delta\int_\Delta$. We denote the triangulation of $\Sigma$ by $T(\epsilon)$. The small parameter $\epsilon$ indicates the ``length'' of the edges of $\Delta$. For each $\Delta$, we single out one of its vertices and called it the base-point $v(\Delta)$ of $\Delta$ and denote its three edges outgoing from $v(\Delta)$ by $s_I(\Delta)$, $I=1,2,3$. Taking the limit $\epsilon\rightarrow0$ corresponds to sharking $\Delta$ to $v(\Delta)$. Let $\alpha_{IJ}(\Delta):=s_I(\Delta)\circ a_{IJ}\circ s^{-1}_J(\Delta)$ be the loop based at $v(\Delta)$, where $a_{IJ}$ is the edge of $\Delta$ from the endpoint of $s_I(\Delta)$ to the endpoint of $s_J(\Delta)$. Then the Euclidean Hamiltonian constraint can be written in the form \cite{Thiemann:2007bk,Thiemann:1996aw}
\begin{align}
H^E(N)&=\frac{1}{2\kappa}\int_{\Sigma}{\rm d}^3x\,N\epsilon_{ijk}\frac{\tilde{E}^a_i\tilde{E}^b_j}{\sqrt{\det(q)}}F^k_{ab}=-\frac{2}{\kappa^2\beta}\sum_{\Delta\in T(\epsilon)}\int_{\Delta}{\rm d}^3x\,\bar{N}\,\tilde{\epsilon}^{abc}{\rm tr}(F_{ab}\{A_c,V\})\notag\\
&=\lim_{\epsilon\rightarrow0}\frac{2}{3\kappa^2\beta}\sum_{\Delta\in T(\epsilon)}\bar{N}(v(\Delta))\epsilon^{IJK}{\rm tr}(h_{\alpha_{IJ}(\Delta)}h_{s_K(\Delta)}\{h_{s_K(\Delta)}^{-1},V\})\notag\\
&=:\lim_{\epsilon\rightarrow0}H^E_{T(\epsilon)}(N)\,,
\end{align}
where $\bar{N}:={\rm sgn}(\det(e_d^l))N$, $A_c:=A^k_c\tau_k$ with $\tau_k$ ($k=1,2,3$) the generators of $su(2)$, $h_{e}\equiv h_e(A)$ is the holonomy  of a connection $A$ along an edge $e$, $V$ denotes the volume function of $\Sigma$, and in the second step we have used the identities
\begin{align}
\epsilon_{ijk}\frac{\tilde{E}^a_i\tilde{E}^b_j}{\sqrt{\det(q)}}&={\rm sgn}(\det(e_d^l))\tilde{\epsilon}^{abc}e^k_c={\rm sgn}(\det(e_d^l))\tilde{\epsilon}^{abc}\frac{2}{\kappa\beta}\{A^l_c,V\}\delta_{kl}, \quad {\rm tr}(\tau_k\tau_l)=-\frac12\delta_{kl}\,.
\end{align}
In what follows we will drop the bar over $N$. Replacing $V$ by $\hat{V}$, holonomies by holonomy operators (since the holonomy operator acts as a multiplication operator, we also omit the hat for simplification of notation),  and the Poisson bracket by $1/(i\hbar)$ times the commutator, the Euclidean Hamiltonian constraint operator reads
\begin{align}
\hat{H}^E_{T(\epsilon)}(N)\label{ham-orign-form}
&=\frac{2}{3i\hbar\kappa^2\beta}\sum_{\Delta\in T(\epsilon)}N(v(\Delta))\epsilon^{IJK}{\rm tr}(h_{\alpha_{IJ}(\Delta)}h_{s_K(\Delta)}[h_{s_K(\Delta)}^{-1},\hat{V}])\notag\\
&=-\frac{2}{3i\hbar\kappa^2\beta}\sum_{\Delta\in T(\epsilon)}N(v(\Delta))\epsilon^{IJK}{\rm tr}(h_{\alpha_{IJ}(\Delta)}h_{s_K(\Delta)}\hat{V}h_{s_K(\Delta)}^{-1})\notag\\
&=:-\frac{2}{3i\hbar\kappa^2\beta}\sum_{\Delta\in T(\epsilon)}N(v(\Delta))\hat{H}^E_{\Delta}\,,
\end{align}
The volume operator $\hat{V}$ actes on a cylindrical function $f_\gamma$ over $\gamma$ by \cite{Ashtekar:1997fb,Thiemann:1996au}
\begin{align}\label{volume-operator}
\hat{V}\cdot\,f_\gamma=\ell_{\rm p}^3\,\beta^{\frac32}\sum_{v\in V(\gamma)}\sqrt{\left|\frac{i}{8\times4}\sum_{I<J<K,\,e_I\cap e_J \cap e_K=v}\varsigma(e_I,e_J,e_K)\;\hat{q}_{IJK}\right|}\,\cdot f_\gamma\,,
\end{align}
where $\ell_{\rm p}\equiv\sqrt{\hbar\kappa}$, $\varsigma(e_I,e_J,e_K)\equiv{\rm sgn}(\det(\dot{e}_I(0),\dot{e}_J(0),\dot{e}_K(0)))$, and
\begin{align}\label{q-IJK}
\hat{q}_{IJK}&:=-4i\epsilon_{ijk}J^i_{e_I}J^j_{e_J}J^k_{e_K}\,,
\end{align}
here $J^i_{e_I}$ is the self-adjoint operator of the right-invariant vector field on the copy of $SU(2)$ corresponding to the $I$-th edge.

It is clear that the operator \eqref{ham-orign-form} depends on the triangulation $T(\epsilon)$. It turns out that the nontrivial action of $\hat{H}^E_{\Delta}$ on a cylindrical function $f_\gamma$ corresponds to $v(\Delta)\cap \gamma\neq\emptyset$, which motivates us to triangulate $\Sigma$ adapted to $\gamma$ \cite{Thiemann:1996aw}. We denote the triangulation adapted to $\gamma$ by $T(\gamma)$. The assignment $T(\gamma)$ which we choose is as follows:\begin{enumerate}
\item[(1)] If $v\in V(\gamma)$ is non-planar with valence bigger than two, a tetrahedron $\Delta$ is assigned to $v$ so that $v(\Delta)=v$ and the three edges starting from $v(\Delta)$ are chosen as the starting segments $(s_I,s_J,s_K)$ of a triple of distinct edges $(e_I,e_J,e_K)$ of $\gamma$ with indecent tangents at $v$.
\item[(2)] If $v$ is coplanar, we add one more edge $e$ incident at $v$ to $\gamma$ such that its tangent at $v$ is transversal to the tangents of edges of $\gamma$ at $v$. The new graph is denoted by $\gamma'$. The three edges of $\Delta$ with base-point $v(\Delta)=v$ come from the starting segment of the new edge $e$ and the other two edges of $\gamma$ with independent tangents incident at $v$.
\end{enumerate}
Then the action of the regularized Euclidean Hamiltonian constraint operator \eqref{ham-orign-form} on $f_\gamma$ reduces to
\begin{align}\label{E-Hamiltonian}
\hat{H}^E_\gamma(N)\cdot  f_\gamma:=\hat{H}^E_{T(\gamma)}(N)\cdot f_\gamma=-\frac{2}{3i\hbar\kappa^2\beta}\sum_{v\in V(\gamma)}N(v)\frac{8}{E(v)}\sum_{v(\Delta)=v}\hat{H}^E_\Delta\cdot  f_\gamma=:\sum_{v\in V(\gamma)}N'_v\hat{H}^E_v\cdot  f_\gamma\,,
\end{align}
where $E(v)$ is the number of non-planar triples of edges of $\gamma$ or $\gamma'$ at $v$, $N'_v:=-\frac{16}{3i\hbar\kappa^2\beta}\frac{N(v)}{E(v)}$, and
\begin{align}\label{H-E-v}
\hat{H}^E_v&:=\sum_{v(\Delta)=v}\epsilon^{IJK}{\rm tr}(h_{\alpha_{IJ}(\Delta)}h_{s_K(\Delta)}\hat{V}h_{s_K(\Delta)}^{-1})\,.
\end{align}
The limit $\epsilon\rightarrow0$ can be taken in the Rovelli-Smolin topology. The label $T$ for the triangulation $T(\gamma)$ can be dropped off since the final limit operator is independent of $\epsilon$.

\subsection{The action of $\hat{H}^E_\gamma(N)$ on a trivalent non-planar vertex of $\gamma$}
The action of the Hamiltonian constraint \eqref{E-Hamiltonian} is local in the sense that it is a sum over independent vertices. Therefore, we can concentrate on its action on a single vertex.  Given a spin network state $T_{\gamma,\vec{j},\vec{i}}(A)$ on a graph $\gamma$, we consider a trivalent non-planar vertex $v\in V(\gamma)$ at which three edges $e_1,e_2,e_3$ incident. The terms in $T_{\gamma,\vec{j},\vec{i}}(A)$ directly associated to $v$ can be represented by \footnote{For calculational convenience, the spin network states considered here are not normalized, although their intertwiners are normalized. They can be normalized by multiplying $\prod_{I=1}^3\sqrt{2j_I+1}$.}
\begin{align}\label{3-valent-snf-algbrac}
T^v_{\gamma,\vec{j},\vec{i}}(A):=\left(i_v\right)_{\,m_1m_2m_3}{[\pi_{j_1}(h_{e_1})]^{m_1}}_{\,n_1}{[\pi_{j_2}(h_{e_2})]^{m_2}}_{\,n_2}{[\pi_{j_3}(h_{e_3})]^{m_3}}_{\,n_3}\,,
\end{align}
where $\left(i_v\right)_{\,m_1m_2m_3}\equiv{\left(i^{\,J=0;\,a_2=j_3}_{j_1,j_2,j_3}\right)_{\,m_1m_2m_3}}^{M=0}$ denotes the intertwiner associated to $v$. The summation in the expression of $\hat{H}^E_v$ in \eqref{H-E-v} is over only one tetrahedron $\Delta$ adapted to $\gamma$ at $v$. We will omit the notation $\Delta$. Then the action of $\hat{H}^E_v$ on $T^v_{\gamma,\vec{j},\vec{i}}(A)$ can be explicitly written as
\begin{align}\label{H-E-three-term}
\hat{H}^E_v\cdot T^v_{\gamma,\vec{j},\vec{i}}(A)&=\epsilon^{IJK}{\rm tr}(h_{\alpha_{IJ}}h_{s_K}\hat{V}h_{s_K}^{-1})\cdot T^v_{\gamma,\vec{j},\vec{i}}(A)=\epsilon^{IJK}{[h_{\alpha_{IJ}}]^A}_B{[h_{s_K}]^B}_C\hat{V}{[h_{s_K}^{-1}]^C}_A\cdot T^v_{\gamma,\vec{j},\vec{i}}(A)\notag\\
&={[h_{\alpha_{23}}-h_{\alpha_{32}}]^A}_B{[h_{s_1}]^B}_C\hat{V}{[h_{s_1}^{-1}]^C}_A\cdot T^v_{\gamma,\vec{j},\vec{i}}(A)+{[h_{\alpha_{31}}-h_{\alpha_{13}}]^A}_B{[h_{s_2}]^B}_C\hat{V}{[h_{s_2}^{-1}]^C}_A\cdot T^v_{\gamma,\vec{j},\vec{i}}(A)\notag\\
&\quad+{[h_{\alpha_{12}}-h_{\alpha_{21}}]^A}_B{[h_{s_3}]^B}_C\hat{V}{[h_{s_3}^{-1}]^C}_A\cdot T^v_{\gamma,\vec{j},\vec{i}}(A)\notag\\
&\equiv\left(\hat{H}^E_{v,s_2s_3s_1}+\hat{H}^E_{v,s_3s_1s_2}+\hat{H}^E_{v,s_1s_2s_3}\right)\cdot T^v_{\gamma,\vec{j},\vec{i}}(A)\,,
\end{align}
where ${[h]^A}_B\equiv{[\pi_{1/2}(h)]^A}_B$. Note that applying $\hat{H}^E_v$ to $T^v_{\gamma,\vec{j},\vec{i}}(A)$ involves the actions of the holonomy and the volume operators.

Before calculating the action \eqref{H-E-three-term}, let us recall the graphical method introduced in the first paper of the series \cite{graph-I}. The $3j$-symbol is represented by
\begin{align}\label{3j-def-graph}
\begin{pmatrix}
j_1 & j_2 & j_3\\
m_1 & m_2 & m_3
\end{pmatrix}
&=\makeSymbol{
\includegraphics[width=2cm]{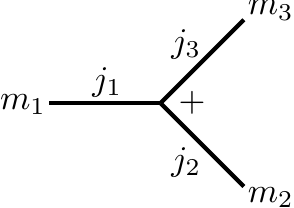}}=\makeSymbol{
\includegraphics[width=2cm]{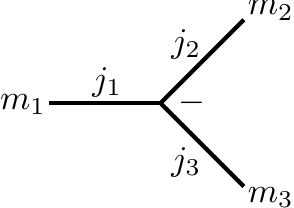}}\,.
\end{align}
The orientation of the node is meant the cyclic order of the lines. A clockwise orientation is denoted by a ``$-$" sign and an anti-clockwise orientation by a ``$+$" sign. Rotation of the diagram does not change the cyclic order of lines, and the angles between two lines as well as their lengths at a node have no significance. Summation over a magnetic quantum number $m$ is graphically represented by joining the free ends of the corresponding lines. The intertwiner associated to $v$ in \eqref{3-valent-snf-algbrac} is represented in graphical formula as (see Appendix A in \cite{graph-I})
\begin{align}\label{intertwiner-original}
\left(i_v\right)_{\,m_1m_2m_3}\equiv{\left(i^{\,J=0;\,a_2=j_3}_{j_1,j_2,j_3}\right)_{\,m_1m_2m_3}}^{M=0}&=\sqrt{d_{j_3}}\makeSymbol{
\includegraphics[width=3cm]{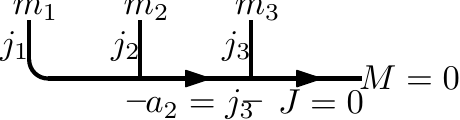}}=\makeSymbol{
\includegraphics[width=2cm]{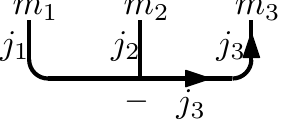}}=(-1)^{2j_3}\makeSymbol{
\includegraphics[width=1.4cm]{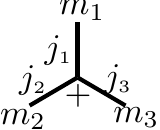}}\,,
\end{align}
where $d_{j_3}\equiv 2j_3+1$, and in the last two steps we have used the following two identities ((A.43) and (A.51) in \cite{graph-I})
\begin{align}
\qquad\makeSymbol{
\includegraphics[width=1.9cm]{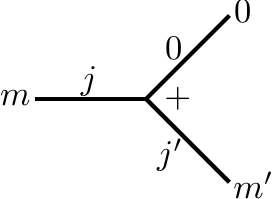}}=\frac{\delta_{j,j'}}{\sqrt{d_j}}\;\makeSymbol{
\includegraphics[width=1.9cm]{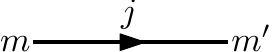}}\label{arrow-3j},\qquad \makeSymbol{
\includegraphics[width=1.9cm]{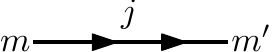}}=(-1)^{2j}\makeSymbol{
\includegraphics[width=1.9cm]{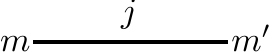}}\,,
\end{align}
here a line with an arrow on it and a line without arrow denote respectively
\begin{align}
C^{(j)}_{m'm}&=C^{mm'}_{(j)}:=(-1)^{j-m}\delta_{m,-m'}=(-1)^{j+m'}\delta_{m,-m'}=\makeSymbol{
\includegraphics[width=1.9cm]{graph/Hamiltonian/wigner-3j-symbol-4}}\,,\qquad
\delta^m_{m'}=\delta_{m,m'}=\makeSymbol{
\includegraphics[width=2cm]{graph/Hamiltonian/wigner-3j-symbol-9}}\,.
\end{align}
The matrix element ${[\pi_j(h_e)]^m}_n$ in spin $j$ representation of the holonomy $h_{e}\equiv h_e(A)$ is denoted by a blue line with an arrow on it as
\begin{align}\label{rep-group-graph}
{[\pi_j(h_e)]^m}_n=\;\makeSymbol{
\includegraphics[width=2.4cm]{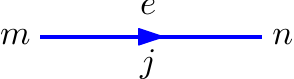}}\,.
\end{align}
The orientation of the arrow is from its row index $m$ to its column index $n$. Then the matrix element ${[\pi_j(h_e^{-1})]^n}_{\,m}$ can be represented by
\begin{align}\label{rep-inverse-graph}
{[\pi_j(h_e^{-1})]^n}_{\,m}={[\pi_j(h_{e^{-1}})]^n}_{\,m}=C^{(j)}_{mm'}\,{[\pi_j(h_e)]^{m'}}_{n'}C^{n'n}_{(j)}=\makeSymbol{
\includegraphics[width=2.4cm]{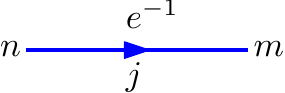}}=\makeSymbol{
\includegraphics[width=3.6cm]{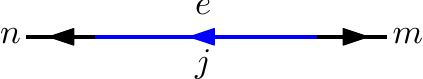}}\,.
\end{align}
Notice that the holonomy acts as multiplication. Hence its action on the spin network states will involve the Clebsch-Gordan series, which can be represented graphically by
\begin{align}\label{holonomy-reps-couple}
\makeSymbol{
\includegraphics[width=2.8cm]{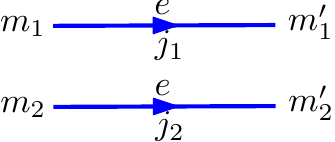}}&=\sum_{j_3}d_{j_3}\makeSymbol{
\includegraphics[width=4.8cm]{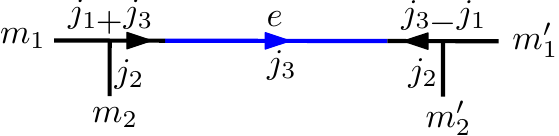}}=\sum_{j_3}d_{j_3}\makeSymbol{
\includegraphics[width=4.8cm]{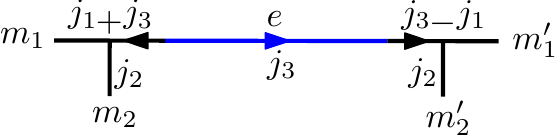}}\,,
\end{align}
and
\begin{align}
\makeSymbol{
\includegraphics[width=2.8cm]{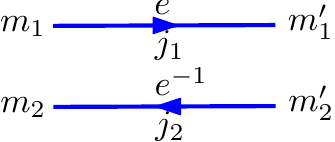}}&=\sum_{j_3}d_{j_3}\makeSymbol{
\includegraphics[width=4.8cm]{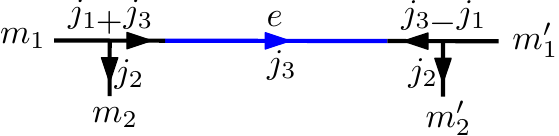}}=\sum_{j_3}d_{j_3}\makeSymbol{
\includegraphics[width=4.8cm]{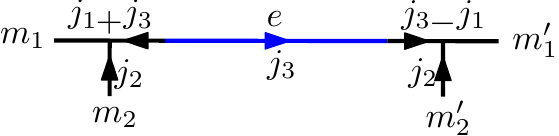}}\label{holonomy-reps-inverse-couple}\,.
\end{align}
With these preparations, the part of $\gamma$ associated at $v$ and the spin network state $T^v_{\gamma,\vec{j},\vec{i}}(A)$ in Eq. \eqref{3-valent-snf-algbrac} can be represented respectively by
\begin{align}\label{graph-snf-origin-edges}
\makeSymbol{
\includegraphics[width=4.6cm]{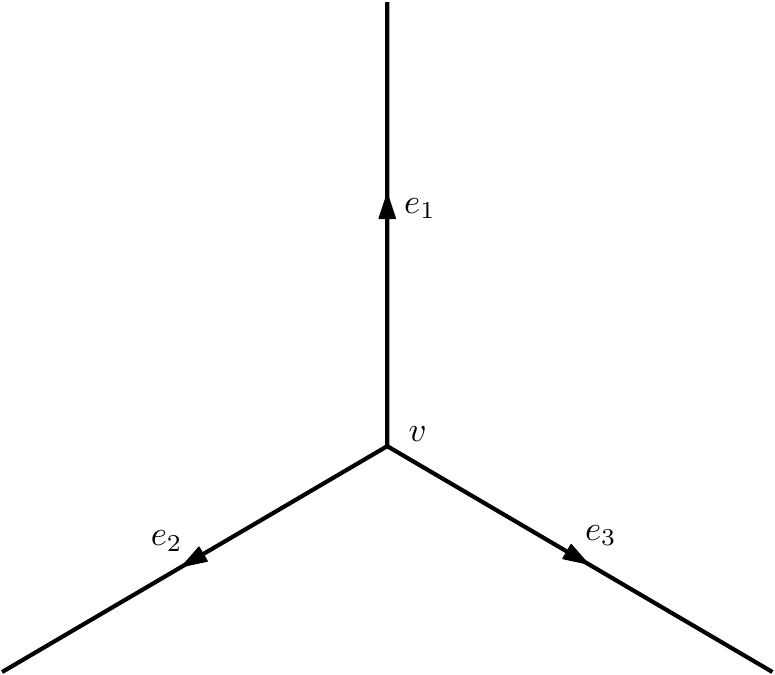}}\qquad\Leftrightarrow\qquad T^v_{\gamma,\vec{j},\vec{i}}\equiv(-1)^{2j_3}\makeSymbol{
\includegraphics[width=4.8cm]{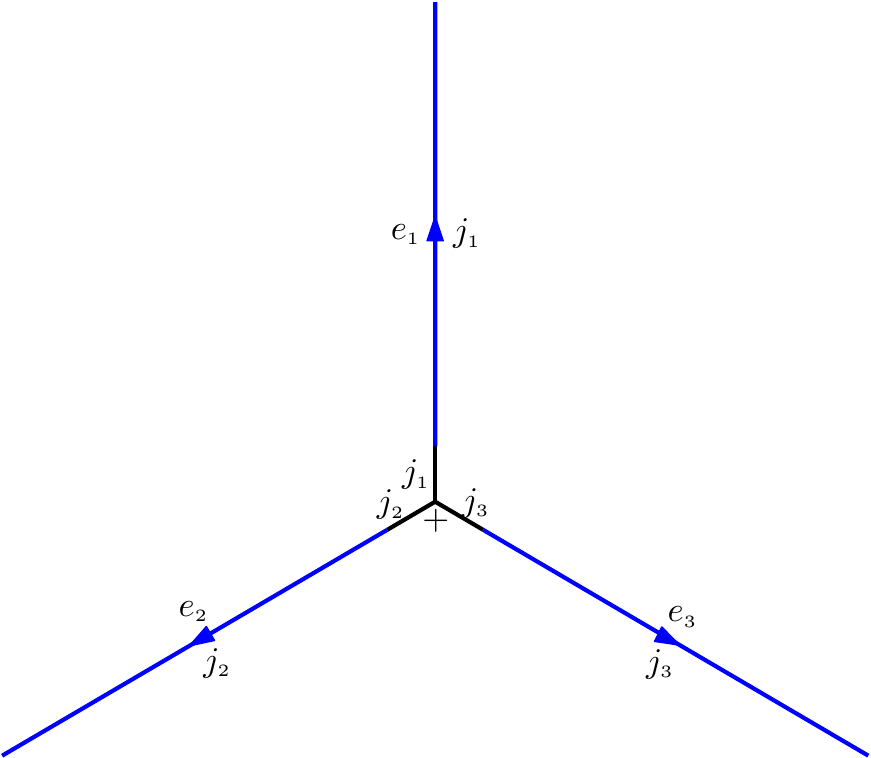}}\,,
\end{align}
where we have omitted the the remaining notations $n_1,n_2,n_3$ in the algebraic form \eqref{3-valent-snf-algbrac}. By introducing three pseudo-vertices $\tilde{v}_I, I=1,2,3$, we subdivide $e_I$ into two parts $s_I$ and $l_I$ such that $e_I=s_I\circ l_I$ and $s_I=s_I(\Delta)$ matching the triangulation $T(\gamma)$. Then $T^v_{\gamma,\vec{j},\vec{i}}(A)$ in Eq. \eqref{3-valent-snf-algbrac} becomes
\begin{align}\label{3-valent-snf-algbrac-1}
T^v_{\gamma,\vec{j},\vec{i}}(A)&=\left(i_v\right)_{\,m_1m_2m_3}{[\pi_{j_1}(h_{s_1})]^{m_1}}_{\,l_1}\delta^{l_1}_{k_1}{[\pi_{j_1}(h_{l_1})]^{k_1}}_{\,n_1}{[\pi_{j_2}(h_{s_2})]^{m_2}}_{\,l_2}\delta^{l_2}_{k_2}{[\pi_{j_2}(h_{l_2})]^{k_2}}_{\,n_2}{[\pi_{j_3}(h_{s_3})]^{m_3}}_{\,l_3}\delta^{l_3}_{k_3}{[\pi_{j_3}(h_{l_3})]^{k_3}}_{\,n_3}\notag\\
&=\left(i_v\right)_{\,m_1m_2m_3}{[\pi_{j_1}(h_{s_1})]^{m_1}}_{\,l_1}{[\pi_{j_2}(h_{s_2})]^{m_2}}_{\,l_2}{[\pi_{j_3}(h_{s_3})]^{m_3}}_{\,l_3}{(i_{\tilde{v}_1})_{k_1}}^{l_1}{(i_{\tilde{v}_2})_{k_2}}^{l_2}{(i_{\tilde{v}_3})_{k_3}}^{l_3}{[\pi_{j_1}(h_{l_1})]^{k_1}}_{\,n_1}{[\pi_{j_2}(h_{l_2})]^{k_2}}_{\,n_2}{[\pi_{j_3}(h_{l_3})]^{k_3}}_{\,n_3}\,,
\end{align}
where ${(i_{\tilde{v}_I})_{k_I}}^{l_I}=\delta_{k_I}^{l_I}$ are the intertwiners associated to $\tilde{v}_I$. Hence the original graph and the corresponding spin network state in \eqref{graph-snf-origin-edges} become
\begin{align}\label{graph-snf-origin}
\makeSymbol{
\includegraphics[width=4.6cm]{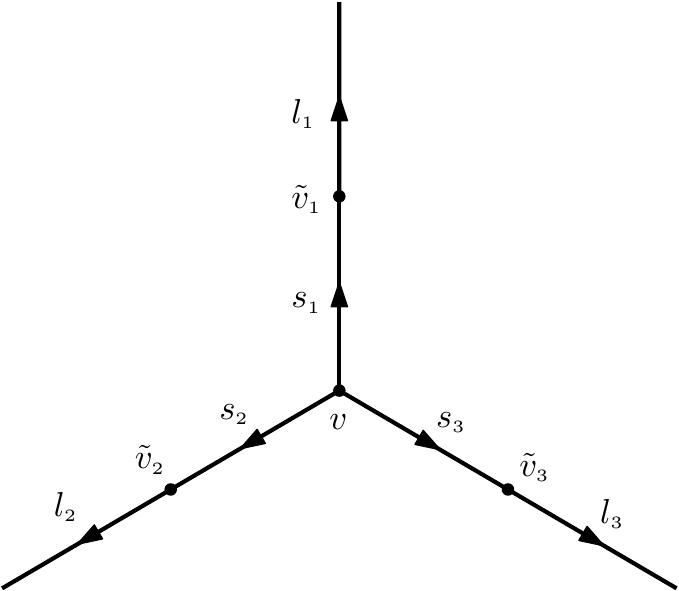}}\qquad\Leftrightarrow\qquad T^v_{\gamma,\vec{j},\vec{i}}(A)\equiv(-1)^{2j_3}\makeSymbol{
\includegraphics[width=5.6cm]{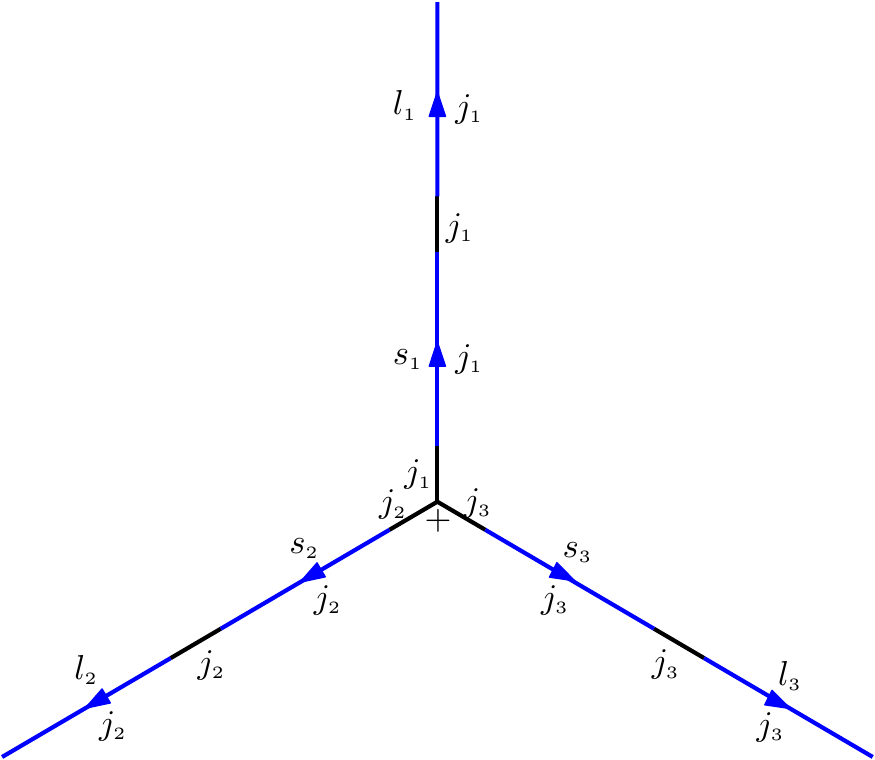}}\,.
\end{align}
We can also single out the corresponding part of $T^v_{\gamma,\vec{j},\vec{i}}(A)$ which only involves the holonomies $h_{s_I}$, and denote it by $T^{v,s}_{\gamma,\vec{j},\vec{i}}(A)$ (the notation $s$ denotes the segments $s_I$), i.e.,
\begin{align}\label{graph-snf-v-s}
T^{v,s}_{\gamma,\vec{j},\vec{i}}(A)&=\left(i_v\right)_{\,m_1m_2m_3}{[\pi_{j_1}(h_{s_1})]^{m_1}}_{\,l_1}{[\pi_{j_2}(h_{s_2})]^{m_2}}_{\,l_2}{[\pi_{j_3}(h_{s_3})]^{m_3}}_{\,l_3}=(-1)^{2j_3}\makeSymbol{
\includegraphics[width=3cm]{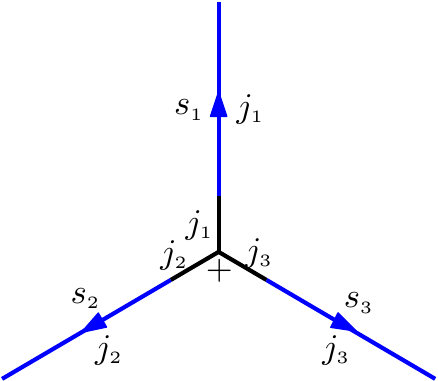}}\,.
\end{align}
Now let us to calculate the action of the first term in \eqref{H-E-three-term} on $T^{v,s}_{\gamma,\vec{j},\vec{i}}(A)$ via the graphical method. The action of ${[h_{s_1}]^B}_C\hat{V}{[h_{s_1}^{-1}]^C}_A$ on $T^{v,s}_{\gamma,\vec{j},\vec{i}}(A)$ can be represented by
\begin{align}\label{action-graph-snf}
{[h_{s_1}]^B}_C\hat{V}{[h_{s_1}^{-1}]^C}_A\left[(-1)^{2j_3}\makeSymbol{
\includegraphics[width=3cm]{graph/Hamiltonian/graph-snf-1}}\right]&={[h_{s_1}]^B}_C\hat{V}\left[(-1)^{2j_3}\makeSymbol{
\includegraphics[width=3cm]{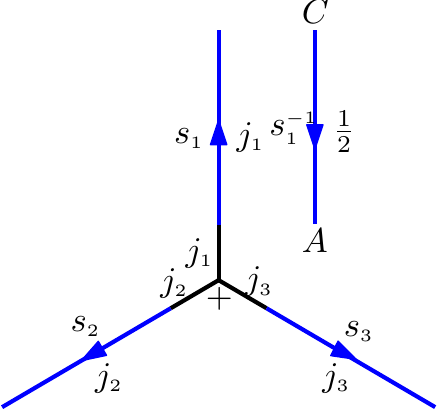}}\right]\notag\\
&={[h_{s_1}]^B}_C\sum_{j'_1}\frac{d_{j'_1}}{\sqrt{d_{j_1}}}\hat{V}\left[(-1)^{2j_3}\sqrt{d_{j_1}}\makeSymbol{
\includegraphics[width=3cm]{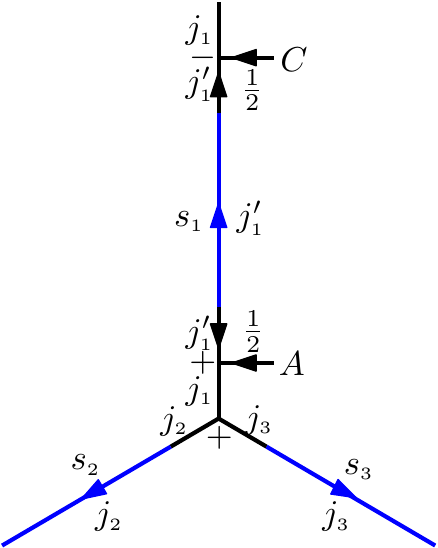}}\right]\,,
\end{align}
where $j'_1=j_1\pm\frac12$, and we have adjusted the coefficients such that the intertwiner $i'_v$ at $v$ is normalized as\begin{align}
(-1)^{2j_3}\sqrt{d_{j_1}}\makeSymbol{
\includegraphics[width=1cm]{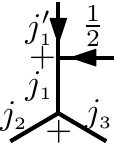}}&=(-1)^{2j_3}\sqrt{d_{j_1}}\makeSymbol{
\includegraphics[width=1cm]{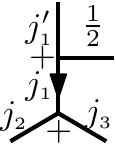}}=\sqrt{d_{j_1}}\makeSymbol{
\includegraphics[width=3cm]{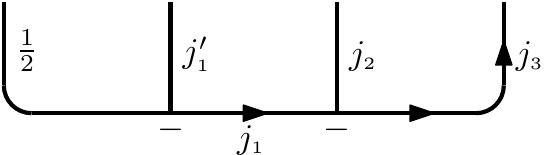}}
=\sqrt{d_{a_2}d_{a_3}}\makeSymbol{
\includegraphics[width=4cm]{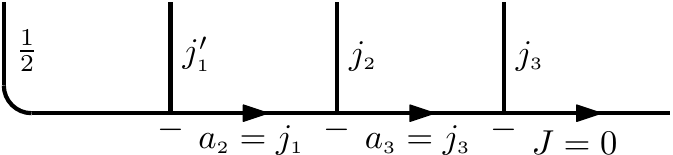}}\,.
\end{align}
Recall that the volume operator \eqref{volume-operator} vanishes coplanar vertices. Hence it has non-trivial action only at $v$, not $\tilde{v}_I$. Its action in \eqref{action-graph-snf} reads
\begin{align}
\hat{V}\left[(-1)^{2j_3}\sqrt{d_{j_1}}\makeSymbol{
\includegraphics[width=1cm]{graph/Hamiltonian/intertwiner-1}}\right]=\frac{\ell_{\rm p}^3\,\beta^{3/2}}{4\sqrt{2}}\sqrt{\left|i\hat{q}_{j'_1j_2j_3}\right|}\left[(-1)^{2j_3}\sqrt{d_{j_1}}\makeSymbol{
\includegraphics[width=1cm]{graph/Hamiltonian/intertwiner-1}}\right]\,,
\end{align}
where the operator $\hat{q}_{j'_1j_2j_3}$ corresponds to the edges $s_1,s_2,s_3$ with spins $j'_1,j_2,j_3$ respectively. Notice that $\hat{q}_{IJK}$ changes only the intermediate momenta $a_I,\cdots, a_{K-1}$ between $j_I$ and $j_K$ of the intertwiner. In our case, the operator $\hat{q}_{j'_1j_2j_3}$ and hence $\hat{V}$ change $a_2=j_1,a_3=j_3$ into
\begin{align}\label{two-combs}
a'_2&=j'_1\pm\frac12=
\begin{cases}
j_1-1,j_1; & \text{for}\;j'_1=j_1-\frac12\\
j_1,j+1; & \text{for}\;j'_1=j_1+\frac12
\end{cases},\qquad\qquad
a'_3=j_3\,.
\end{align}
Hence we have
\begin{align}\label{volume-action-4-dim}
&\hat{V}\left[(-1)^{2j_3}\sqrt{d_{j_1}}\makeSymbol{
\includegraphics[width=1cm]{graph/Hamiltonian/intertwiner-1}}\right]=\hat{V}
\left[\sqrt{d_{a_2}d_{a_3}}\makeSymbol{
\includegraphics[width=4cm]{graph/Hamiltonian/intertwiner-4}}\right]
=\sum_{a'_2,a'_3}\langle a'_2,a'_3|\hat{V}|a_2,a_3 \rangle\left[\sqrt{d_{a'_2}d_{a'_3}}\makeSymbol{
\includegraphics[width=4cm]{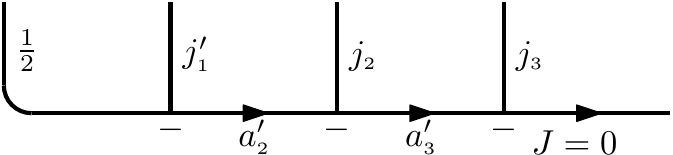}}\right]\,,
\end{align}
where $|a'_2,a'_3\rangle$ and $|a_2,a_3\rangle$ are the simplified forms of the intertwiners corresponding to two different chosen intermediate momenta for the fixed anglular momenta $\frac12,j'_1,j_2,j_3$ and the resulting $J=0$. For given values of four spins $\frac12,j'_1,j_2,j_3$, there are two allowed combinations of intermediate momenta \eqref{two-combs}. Hence the corresponding intertwiner space associated to $v$ has dimension 2. Furthermore, the volume operator is automatically diagonal on the 2-dimensional intertwiner space. This fact was pointed out in \cite{Thiemann:1996at,Borissov:1997ji}, which will also be presented in \ref{volume-eigenvalue}. Therefore we obtain
\begin{align}
\hat{V}\,\left[(-1)^{2j_3}\sqrt{d_{j_1}}\makeSymbol{
\includegraphics[width=1cm]{graph/Hamiltonian/intertwiner-1}}\right]&=V(j'_1,j_2,j_3)\,\left[(-1)^{2j_3}\sqrt{d_{j_1}}\makeSymbol{
\includegraphics[width=1cm]{graph/Hamiltonian/intertwiner-1}}\right]\,,
\end{align}
or
\begin{align}
\hat{V}\,\left[\sqrt{d_{j_1}}\makeSymbol{
\includegraphics[width=1cm]{graph/Hamiltonian/intertwiner-1}}\right]&=V(j'_1,j_2,j_3)\,\left[\sqrt{d_{j_1}}\makeSymbol{
\includegraphics[width=1cm]{graph/Hamiltonian/intertwiner-1}}\right]\,,
\end{align}
with
\begin{align}
V(j'_1,j_2,j_3)&\equiv V(1/2,j'_1,j_2,j_3;a_2=j'_1+1/2,a_3=j_3)\notag\\
&\equiv\frac{\ell_{\rm p}^3\,\beta^{3/2}}{4\sqrt{2}}\left[(j'_1+j_2+j_3+\frac32)(j'_1+j_2-j_3+\frac12)(j'_1-j_2+j_3+\frac12)(-j'_1+j_2+j_3+\frac12)\right]^{\frac14}\,.
\end{align}
Hence we have
\begin{align}\label{snf-after-action}
&{[h_{s_1}]^B}_C\hat{V}{[h_{s_1}^{-1}]^C}_A\left[(-1)^{2j_3}\makeSymbol{
\includegraphics[width=3cm]{graph/Hamiltonian/graph-snf-1}}\right]={[h_{s_1}]^B}_C\sum_{j'_1}V(j'_1,j_2,j_3)(-1)^{2j_3}d_{j'_1}\makeSymbol{
\includegraphics[width=3cm]{graph/Hamiltonian/graph-snf-3}}\notag\\
&=\sum_{j'_1}V(j'_1,j_2,j_3)(-1)^{2j_3}d_{j'_1}\makeSymbol{
\includegraphics[width=3cm]{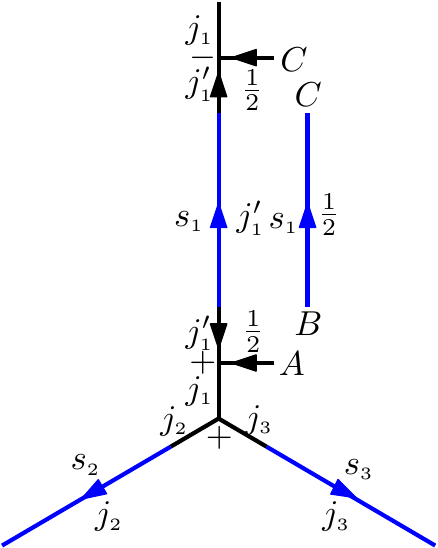}}
=\sum_{j'_1}V(j'_1,j_2,j_3)(-1)^{2j_3}d_{j'_1}\sum_{j''_1}d_{j''_1}\makeSymbol{
\includegraphics[width=3cm]{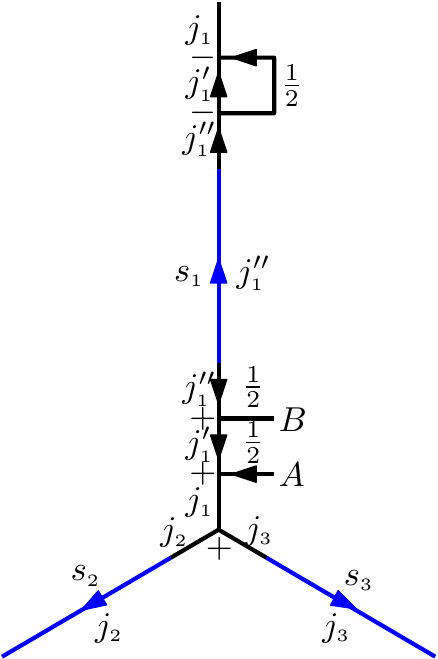}}\notag\\
&=\sum_{j'_1}V(j'_1,j_2,j_3)(-1)^{2j_3}d_{j'_1}\sum_{j''_1}d_{j''_1}\makeSymbol{
\includegraphics[width=3cm]{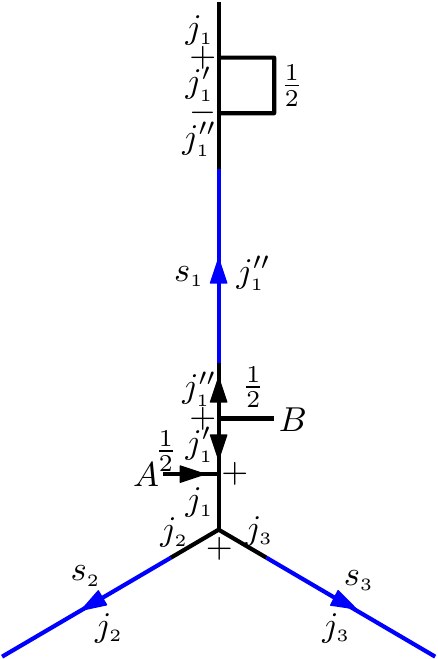}}=\sum_{j'_1}V(j'_1,j_2,j_3)(-1)^{2j_3}d_{j'_1}\sum_{j''_1}\frac{d_{j''_1}\delta_{j_1,j''_1}}{d_{j_1}}\makeSymbol{
\includegraphics[width=3cm]{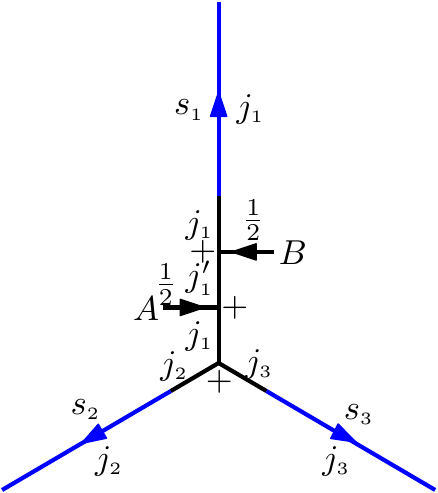}}\notag\\
&=\sum_{j'_1}V(j'_1,j_2,j_3)(-1)^{2j_3}d_{j'_1}\makeSymbol{
\includegraphics[width=3cm]{graph/Hamiltonian/graph-snf-7}}\,,
\end{align}
where in the fourth step we have changed the orientation of two arrows with spin $j'_1$ by the rule \eqref{arrow-flip}, and then used the rule \eqref{three-arrow-adding} to remove three arrows with the same orientation joint with a $3j$-symbol, and we also used the rule ((A.54) in \cite{graph-I})
\begin{align}\label{loop-id}
\makeSymbol{
\includegraphics[width=3cm]{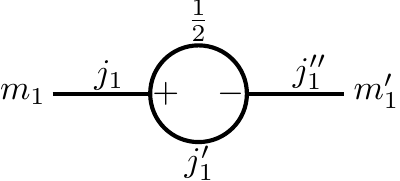}}&=\frac{\delta_{j_1,j_1''}}{d_{j_1}}\,
\makeSymbol{
\includegraphics[width=2.4cm]{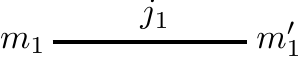}}\,,
\end{align}
to remove a loop in the fifth step. Thus the action of $\hat{H}^E_{v,s_2s_3s_1}={[h_{\alpha_{23}}-h_{\alpha_{32}}]^A}_B{[h_{s_1}]^B}_C\hat{V}{[h_{s_1}^{-1}]^C}_A$ on $T^{v,s}_{\gamma,\vec{j},\vec{i}}(A)$ is given by
\begin{align}\label{H-231}
&\hat{H}^E_{v,s_2s_3s_1}\left[(-1)^{2j_3}\makeSymbol{
\includegraphics[width=3cm]{graph/Hamiltonian/graph-snf-1}}\right]\notag\\
&=\sum_{j'_1}V(j'_1,j_2,j_3)(-1)^{2j_3}d_{j'_1}\left[\makeSymbol{
\includegraphics[width=3cm]{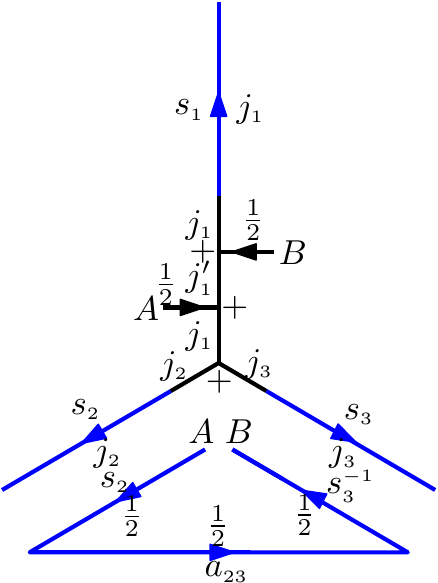}}-\makeSymbol{
\includegraphics[width=3cm]{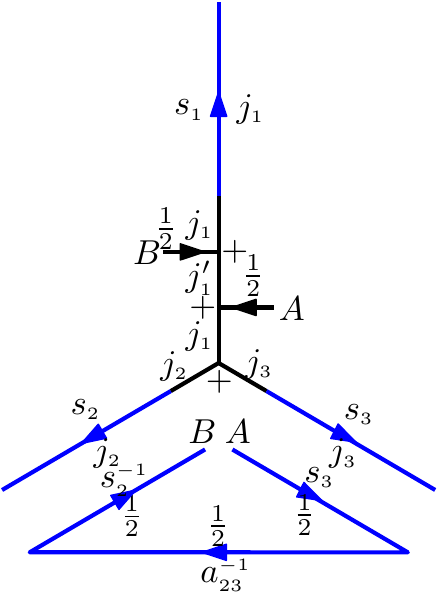}}\right]\notag\\
&=\sum_{j'_1}V(j'_1,j_2,j_3)(-1)^{2j_3}d_{j'_1}\left[\makeSymbol{
\includegraphics[width=3cm]{graph/Hamiltonian/graph-snf-loop-1}}-\makeSymbol{
\includegraphics[width=3cm]{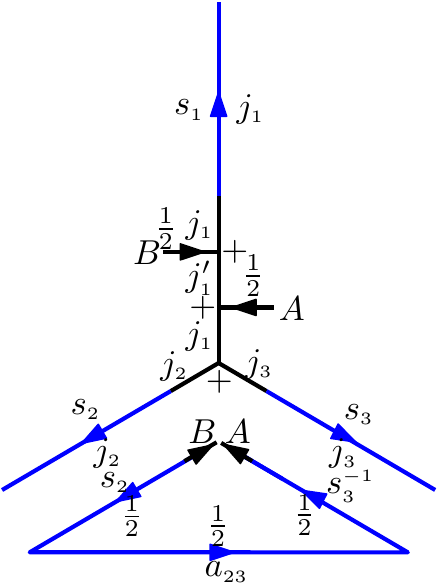}}\right]\notag\\
&=\sum_{j'_1}V(j'_1,j_2,j_3)(-1)^{2j_3}d_{j'_1}
\sum_{j'_2,j'_3}d_{j'_2}d_{j'_3}\left[\makeSymbol{
\includegraphics[width=4.4cm]{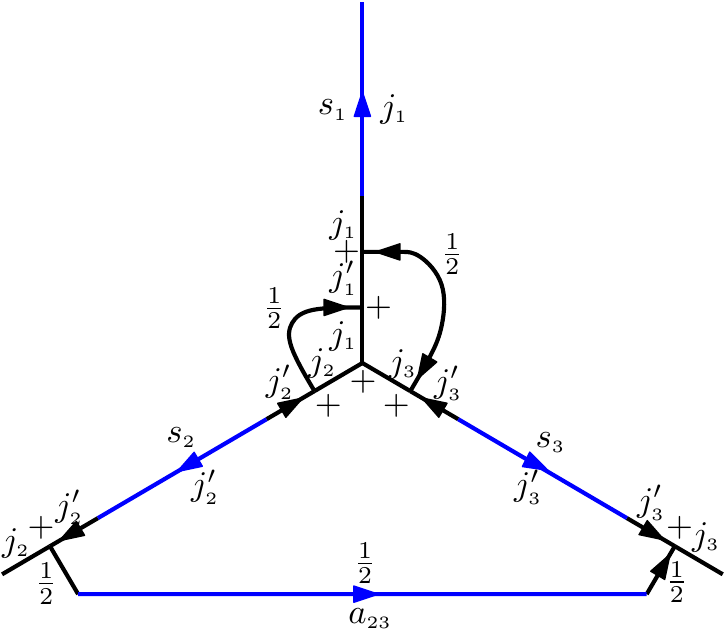}}-\makeSymbol{
\includegraphics[width=4.4cm]{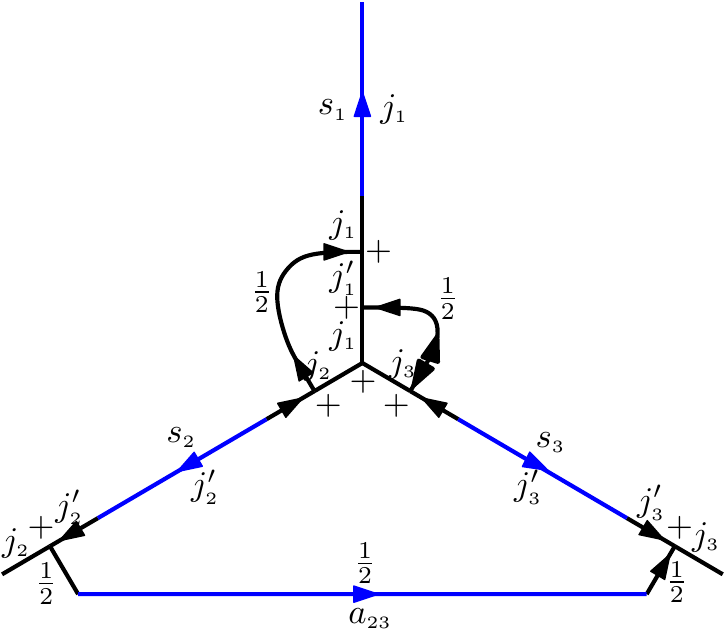}}\right]\notag\\
&=\sum_{j'_1,j'_2,j'_3}V(j'_1,j_2,j_3)(-1)^{2j_3}d_{j'_1}d_{j'_2}d_{j'_3}\notag\\
&\qquad\times\left[-(-1)^{j_1+j'_1+\frac12}(-1)^{j_3+j'_3+\frac12}
\begin{Bmatrix}
j_1 & \frac12 & j'_1\\
j'_2 & j_3 & j_2
\end{Bmatrix}
\begin{Bmatrix}
j_1 & \frac12 & j'_1\\
j_3 & j'_2 & j'_3
\end{Bmatrix}-
(-1)^{j_1-j'_1+\frac12}(-1)^{j_2+j'_2+\frac12}
\begin{Bmatrix}
j_1 & \frac12 & j'_1\\
j_2 & j'_3 & j'_2
\end{Bmatrix}
\begin{Bmatrix}
j_1 & \frac12 & j'_1\\
j'_3 & j_2 & j_3
\end{Bmatrix}
\right]\notag\\
&\qquad\times\makeSymbol{
\includegraphics[width=4.4cm]{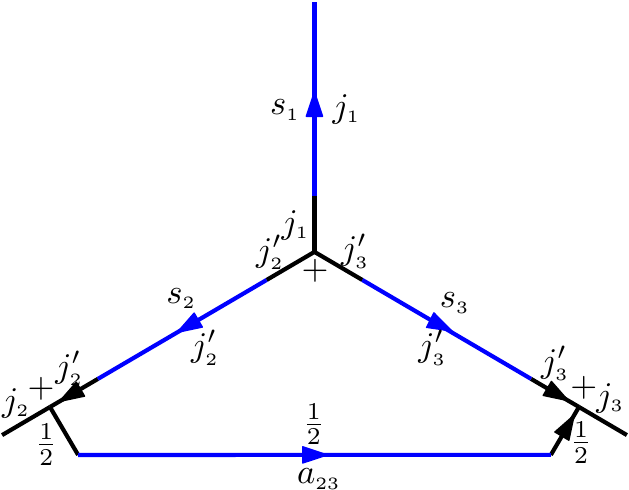}}\,,
\end{align}
where in the second step we have used \eqref{rep-inverse-graph}, and in the last step we have used the two identities (see \ref{Appendix-identity-graph-proof} for proof)
\begin{align}\label{ham-6j-1}
\makeSymbol{
\includegraphics[width=1.5cm]{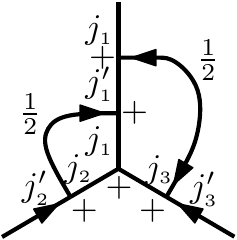}}&=-(-1)^{j_1+j'_1+\frac12}(-1)^{j_3+j'_3+\frac12}
\begin{Bmatrix}
j_1 & \frac12 & j'_1\\
j'_2 & j_3 & j_2
\end{Bmatrix}
\begin{Bmatrix}
j_1 & \frac12 & j'_1\\
j_3 & j'_2 & j'_3
\end{Bmatrix}\makeSymbol{
\includegraphics[width=1cm]{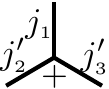}}\,,\\
\makeSymbol{
\includegraphics[width=1.5cm]{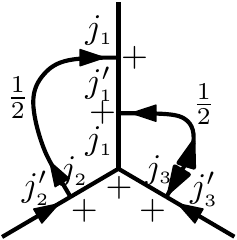}}&=(-1)^{j_1-j'_1+\frac12}(-1)^{j_2+j'_2+\frac12}
\begin{Bmatrix}
j_1 & \frac12 & j'_1\\
j_2 & j'_3 & j'_2
\end{Bmatrix}
\begin{Bmatrix}
j_1 & \frac12 & j'_1\\
j'_3 & j_2 & j_3
\end{Bmatrix}\makeSymbol{
\includegraphics[width=1cm]{graph/Hamiltonian/graph-origin-snf-loop-id-1-8}}\label{ham-6j-2}\,.
\end{align}
Eq. \eqref{H-231} enables us to write directly down the results for $(I,J,K)\in\{(2,3,1),(3,1,2),(1,2,3)\}$ as
\begin{align}\label{H-IJK}
&\hat{H}^E_{v,s_Is_Js_K}\makeSymbol{
\includegraphics[width=3cm]{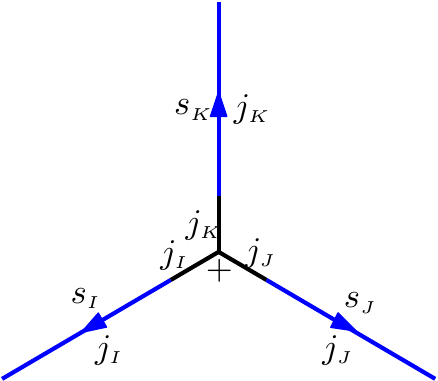}}=\sum_{j'_I,j'_J}[-H(j'_I,j'_J,j_K)]\makeSymbol{
\includegraphics[width=4.4cm]{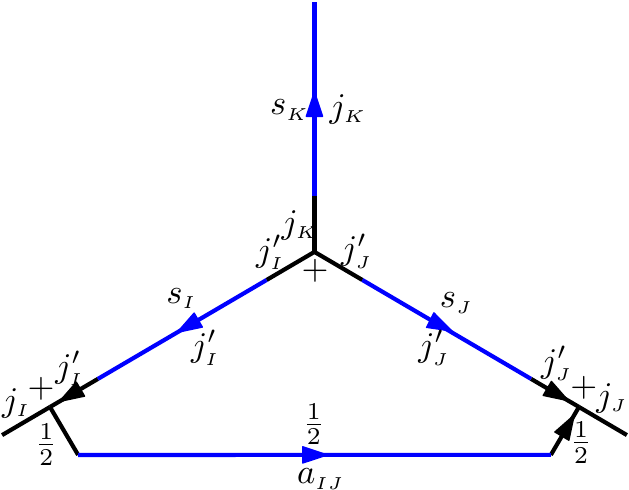}}\,,
\end{align}
where
\begin{align}
H(j'_I,j'_J,j_K)&=\sum_{j'_K}V(j'_K,j_I,j_J)d_{j'_K}d_{j'_I}d_{j'_J}\notag\\
&\hspace{0.5cm}\times\left[(-1)^{j_K+j'_K+\frac12}(-1)^{j_J+j'_J+\frac12}
\begin{Bmatrix}
j_K & \frac12 & j'_K\\
j'_I & j_J & j_I
\end{Bmatrix}
\begin{Bmatrix}
j_K & \frac12 & j'_K\\
j_J & j'_I & j'_J
\end{Bmatrix}+
(-1)^{j_K-j'_K+\frac12}(-1)^{j_I+j'_I+\frac12}
\begin{Bmatrix}
j_K & \frac12 & j'_K\\
j_I & j'_J & j'_I
\end{Bmatrix}
\begin{Bmatrix}
j_K & \frac12 & j'_K\\
j'_J & j_I & j_J
\end{Bmatrix}
\right]\,.
\end{align}
Taking account of the identity $(-1)^{2j_3+1}=(-1)^{2j'_3}$, the action of $\hat{H}^E_v$ on $T^{v,s}_{\gamma,\vec{j},\vec{i}}(A)$ can be explicitly written down as
\begin{align}\label{action-H-E-v}
\hat{H}^E_v\left[(-1)^{2j_3}\makeSymbol{
\includegraphics[width=3cm]{graph/Hamiltonian/graph-snf-1}}\right]&=\sum_{j'_2,j'_3}H(j'_2,j'_3,j_1)\left[(-1)^{2j'_3}\makeSymbol{
\includegraphics[width=4.4cm]{graph/Hamiltonian/graph-snf-loop-6}}\right]\notag\\
&\quad+\sum_{j'_3,j'_1}H(j'_3,j'_1,j_2)\left[(-1)^{2j'_3}\makeSymbol{
\includegraphics[width=3.6cm]{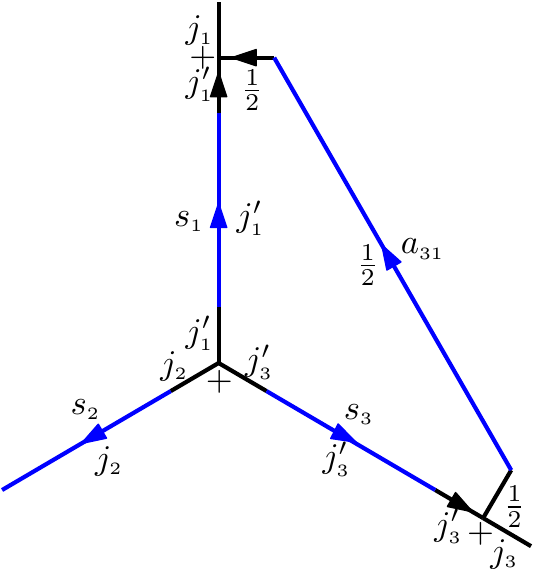}}\right]\notag\\
&\quad-\sum_{j'_1,j'_2}H(j'_1,j'_2,j_3)\left[(-1)^{2j_3}\makeSymbol{
\includegraphics[width=3.6cm]{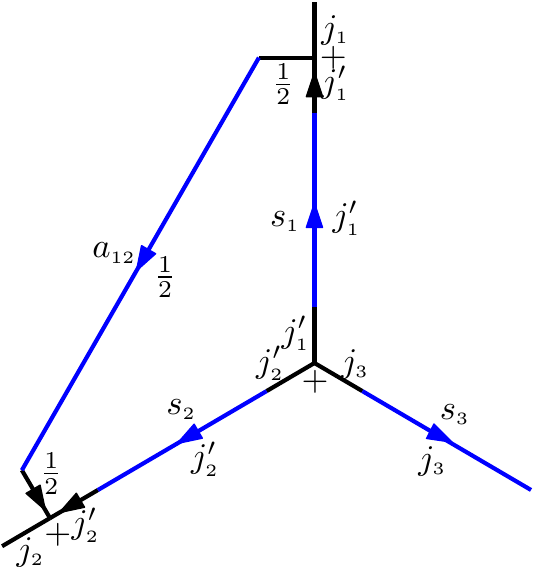}}\right]\,.
\end{align}

\section{The inverse volume operator}\label{sec-III}
The Hamiltonian of a massless scalar field reads
\begin{align}\label{H-phi-all}
H_\phi(N)&=\frac12\int {\rm d}^3x\,N(x)\left[\frac{\pi^2}{\sqrt{\det(q)}}+\sqrt{\det(q)}\,q^{ab}(\partial_a\phi)\partial_b\phi\right](x)\equiv\frac12\left[H_{{\rm kin},\phi}(N)+H_{{\rm der},\phi}(N)\right]\,,
\end{align}
where $\pi$ is the momentum conjugate to $\phi$. In order to quantize $H_{\phi}(N)$ in framework of the scalar field coupled to gravity in LQG, the term $H_{{\rm kin},\phi}(N)$ can be regularized as \cite{{Thiemann:1997rt}}
\begin{align}\label{H-phi-1}
H_{{\rm kin},\phi}(N)&=\int_{\Sigma}{\rm d}^3x\,N(x)\frac{\pi^2(x)}{\sqrt{\det(q)(x)}}\notag\\
&=\lim_{\epsilon\rightarrow0}\int_{\Sigma}{\rm d}^3x\,N(x)\,\pi(x)\int_{\Sigma}{\rm d}^3y\,\pi(y)\int_{\Sigma}{\rm d}^3u\frac{\det(e^i_a)}{\left[\epsilon^3\sqrt{\det(q)}\right]^{3/2}}(u)\int_{\Sigma}{\rm d}^3w\frac{\det(e^i_a)}{\left[\epsilon^3\sqrt{\det(q)}\right]^{3/2}}(w)\;\chi_\epsilon(x,y)\chi_\epsilon(x,u)\chi_\epsilon(x,w)\notag\\
&=\frac{2^6\cdot 2^6}{3!\cdot 3!\cdot\kappa^6\cdot\beta^6}\lim_{\epsilon\rightarrow0}\int_{\Sigma}{\rm d}^3x\,N(x)\,\pi(x)\int_{\Sigma}{\rm d}^3y\,\pi(y)\int_{\Sigma}{\rm d}^3u\;\tilde{\epsilon}^{abc}\epsilon_{ijk}\{A^i_a(u),V(u,\epsilon)^{\frac12}\}\{A^j_b(u),V(u,\epsilon)^{\frac12}\}\{A^k_c(u),V(u,\epsilon)^{\frac12}\}\notag\\
&\hspace{1.5cm}\times\int_{\Sigma}{\rm d}^3w\;\tilde{\epsilon}^{def}\epsilon_{lmn}\{A^l_d(w),V(w,\epsilon)^{\frac12}\}\{A^m_e(w),V(w,\epsilon)^{\frac12}\}\{A^n_f(w),V(w,\epsilon)^{\frac12}\}\times\;\chi_\epsilon(x,y)\chi_\epsilon(x,u)\chi_\epsilon(x,w)\,,
\end{align}
where we have inserted $1=[\det(e^i_a)]^2/\left[\sqrt{\det(q)}\right]^2$ in the second step, used $e^i_a(x)=\frac{2}{\kappa\beta}\{A^i_a(x),V(x,\epsilon)\}$ and absorbed $V(x,\epsilon):=\epsilon^3\sqrt{\det(q)}(x)$ in the denominator into the Poisson bracket in the last step. Again we introduce a triangulation $T(\gamma)$ of $\Sigma$ adapted to a graph $\gamma$. For a given tetrahedron $\Delta$ and its edge $s_I(\Delta)=:s_I$, by the identity
\begin{align}
\int_{s_I(\Delta)}{\rm d}^3x\,\{A^i_a(x),V(x,\epsilon)^{\frac12}\}&=2{\rm tr}\left(\tau_ih_I\{h_I^{-1},V(v,\epsilon)^{\frac12}\}\right)+o(\epsilon^2)\,,\qquad h_I\equiv h_{s_I(\Delta)}\,,
\end{align}
Eq. \eqref{H-phi-1} can be reduced to
\begin{align}\label{H-phi-2}
H_{{\rm kin},\phi}(N)&=\frac{2^{22}}{3^2\cdot\kappa^6\cdot\beta^6}\lim_{\epsilon\rightarrow0}\int_{\Sigma}{\rm d}^3x\,N(x)\,\pi(x)\int_{\Sigma}{\rm d}^3y\,\pi(y)\notag\\
&\hspace{1cm}\times\sum_{v,v'\in V(\gamma)}\frac{1}{E(v)E(v')}\sum_{\substack{s_I\cap s_J\cap s_K=v\\s_L\cap s_M\cap s_N=v'}}\epsilon^{IJK}\epsilon^{LMN}\epsilon^{ijk}\epsilon^{lmn}{\rm tr}\left(\tau_ih_I\{h_I^{-1},V(v,\epsilon)^{\frac12}\}\right){\rm tr}\left(\tau_lh_L\{h_L^{-1},V(v,\epsilon)^{\frac12}\}\right)\notag\\
&\hspace{2cm}\times{\rm tr}\left(\tau_jh_J\{h_J^{-1},V(v,\epsilon)^{\frac12}\}\right){\rm tr}\left(\tau_mh_M\{h_M^{-1},V(v',\epsilon)^{\frac12}\}\right){\rm tr}\left(\tau_kh_K\{h_K^{-1},V(v',\epsilon)^{\frac12}\}\right){\rm tr}\left(\tau_nh_N\{h_N^{-1},V(v',\epsilon)^{\frac12}\}\right)\notag\\
&\hspace{2cm}\times\;\chi_\epsilon(x,y)\chi_\epsilon(x,v)\chi_\epsilon(x,v')\,.
\end{align}
Replacing $\pi$ by $-i\hbar\kappa\delta/\delta\phi$, Poisson brackets by commutators times $1/(i\hbar)$, and substituting $V\rightarrow\hat{V}$, $H_{{\rm kin},\phi}(N)$ can be quantized as
\begin{align}\label{H-phi-op-1}
\hat{H}_{{\rm kin},\phi}(N)_{\gamma}&=\frac{(-i)^22^{22}}{i^63^2\hbar^4\kappa^4\beta^6}\lim_{\epsilon\rightarrow0}\sum_{v,v',v'',v'''\in V(\gamma)}N(v'')X(v'')X(v''')\;\chi_\epsilon(v'',v''')\chi_\epsilon(v'',v)\chi_\epsilon(v'',v')\notag\\
&\hspace{2cm}\times\frac{1}{E(v)E(v')}\sum_{\substack{s_I\cap s_J\cap s_K=v\\s_L\cap s_M\cap s_N=v'}}\epsilon^{IJK}\epsilon^{LMN}\epsilon_{ijk}\epsilon_{lmn}\,{}^{(\frac12)}\!\hat{e}^i_I(v){}^{(\frac12)}\!\hat{e}^l_L(v'){}^{(\frac12)}\!\hat{e}^j_J(v){}^{(\frac12)}\!\hat{e}^m_M(v'){}^{(\frac12)}\!\hat{e}^k_K(v){}^{(\frac12)}\!\hat{e}^n_N(v')\,,
\end{align}
where $X(v):=\frac12\left[X_R(v)+X_L(v)\right]$ is the sum of left and right invariant vector fields acting on the point holomomies $U(v)$ defined in \cite{Thiemann:1997rq}, and
\begin{align}
{}^{(\frac12)}\!\hat{e}^i_I(v):={\rm tr}\left(\tau_ih_I[h_I^{-1},\hat{V}^\frac12]\right)=-{\rm tr}\left(\tau_ih_I\hat{V}^\frac12\,h_I^{-1}\right)
\end{align}
is $\epsilon$-independent for sufficiently small $\epsilon$. For sufficiently small $\epsilon$, the three characteristic functions in \eqref{H-phi-op-1} vanish unless $v=v'=v''=v'''$. Taking the limit $\epsilon\rightarrow0$ yields
\begin{align}\label{H-phi-op-2}
\hat{H}_{{\rm kin},\phi}(N)_{\gamma}&=\frac{2^{23}}{3\hbar^4\kappa^4\beta^6}\sum_{v\in V(\gamma)}\frac{N(v)}{E(v)^2}X(v)X(v)\sum_{\substack{s_I\cap s_J\cap s_K=v\\s_L\cap s_M\cap s_N=v}}\epsilon^{IJK}\epsilon^{LMN}\,\delta_{il}{}^{(\frac12)}\!\hat{e}^i_I(v){}^{(\frac12)}\!\hat{e}^l_L(v)\delta_{jm}{}^{(\frac12)}\!\hat{e}^j_J(v){}^{(\frac12)}\!\hat{e}^m_M(v)\delta_{kn}{}^{(\frac12)}\!\hat{e}^k_K(v){}^{(\frac12)}\!\hat{e}^n_N(v)\notag\\
&=:\frac{2^{23}}{3\hbar^4\kappa^4\beta^6}\sum_{v\in V(\gamma)}\frac{N(v)}{E(v)^2}X(v)X(v)\widehat{V^{-1}}_{{\rm alt},v}\,,
\end{align}
where we have used $\epsilon_{ijk}\epsilon_{lmn}=3!\delta^i_{[l}\delta^j_m\delta^k_{n]}$.
The operator $\widehat{V^{-1}}_{{\rm alt},v}$ is the quantum version of $\frac{1}{\epsilon^3\sqrt{\det(q)}(x)}=\frac{1}{V(x,\epsilon)}$ up to a constant, and thus it is called the {\em inverse volume operator}. By introducing the manifestly gauge invariant operators \cite{Brunnemann:2005ip}
\begin{align}\label{q-IJ}
\hat{q}_{IJ}(v):=\delta_{ij}{}^{(\frac12)}\!\hat{e}^i_I(v){}^{(\frac12)}\!\hat{e}^j_J(v)\,,
\end{align}
the inverse volume operator $\widehat{V^{-1}}_{{\rm alt},v}$ can be represented in terms of $\hat{q}_{IJ}(v)$ as
\begin{align}\label{inverse-volume-operator}
\widehat{V^{-1}}_{{\rm alt},v}\cdot f_\gamma=\sum_{\substack{s_I\cap s_J\cap s_K=v\\s_L\cap s_M\cap s_N=v}}\epsilon^{IJK}\epsilon^{LMN}\hat{q}_{IL}(v)\hat{q}_{JM}(v)\hat{q}_{KN}(v)\cdot f_\gamma\,.
\end{align}
For the convenience of graphical calculus, one usually introduces the spherical tensors (or the irreducible tensor operators) $\tau_\mu$ ($\mu=0,\pm1$) corresponding to $\tau_i$ ($i=1,2,3$) by
\begin{align}\label{tau-mu}
\tau_0:=\tau_3\,,\qquad \tau_{\pm1}:=\mp\frac{1}{\sqrt{2}}\left(\tau_1\pm i \tau_2\right)\,.
\end{align}
Then $\hat{q}_{IJ}(v)$ can be represented in terms of $\tau_\mu$ as
\begin{align}
\hat{q}_{IJ}(v)&=\delta^{ij}{\rm tr}\left(\tau_ih_I\hat{V}^\frac12\,h_I^{-1}\right){\rm tr}\left(\tau_jh_J\hat{V}^\frac12\,h_J^{-1}\right)=-{\rm tr}\left(\tau_{\mu'} h_I\hat{V}^\frac12\,h_I^{-1}\right)C^{\mu'\mu}_{(1)}{\rm tr}\left(\tau_{\mu}h_J\hat{V}^\frac12\,h_J^{-1}\right)\notag\\
&=-\left[{\rm tr}\left(\tau_\mu h_I\hat{V}^\frac12\,h_I^{-1}\right)\right]^\dag{\rm tr}\left(\tau_\mu h_J\hat{V}^\frac12\,h_J^{-1}\right)\notag\\
&=:-\left[{}^{(\frac12)}\!\hat{e}^\mu_I(v)\right]^\dag {}^{(\frac12)}\!\hat{e}^\mu_J(v)\,,
\end{align}
where we have used the following identity in the second step (see Appendix B.1 in \cite{graph-I} for proof)
\begin{align}\label{two-tau-relation}
{[\pi_{j_I}(\tau_i)]^{n_I}}_{m_I}\delta^{ij}{[\pi_{j_J}(\tau_j)]^{n_J}}_{m_J}
&=-{[\pi_{j_I}(\tau_{\mu'})]^{n_I}}_{m_I}C^{\mu'\mu}_{(1)}{[\pi_{j_J}(\tau_{\mu}
)]^{n_J}}_{m_J}\,,\qquad C^{\mu'\mu}_{(1)}=C^{\mu\mu'}_{(1)}\equiv(-1)^{1+\mu}\delta_{\mu,-\mu'}\,,
\end{align}
and the following identities in the third step
\begin{align}
\overline{{(h_I)^A}_B}={(h_I^{-1})^B}_A\,,\qquad \overline{{(\tau_i)^A}_B}=-{(\tau_i)^B}_A\,, \qquad \overline{{(\tau_\mu)^A}_B}={(\tau_{\mu'})^B}_AC^{\mu'\mu}_{(1)}\,,
\end{align}
here the overline denotes complex conjugation.

In what follows, we consider the action of $\widehat{V^{-1}}_{{\rm alt},v}$ on $T^{v,s}_{\gamma,\vec{j},\vec{i}}(A)$ at a trivalent non-planar vertex $v$. Notice that the intertwiner space associated to $v$, which will be acted by $\widehat{V^{-1}}_{{\rm alt},v}$, is of one dimension. Hence the gauge-invariant operators $\hat{q}_{IJ}(v)$ and $\widehat{V^{-1}}_{{\rm alt},v}$
take eigenvalues on the orthonormal spin network states of $T^{v,s}_{\gamma,\vec{j},\vec{i}}(A)$,
\begin{align}\label{normal-snf-v}
T^{v,s,{\rm norm}}_{\gamma,\vec{j},\vec{i}}(A):=\sqrt{d_{j_1}d_{j_2}d_{j_3}}\;T^{v,s}_{\gamma,\vec{j},\vec{i}}(A)\,.
\end{align}
Therefore we have
\begin{align}
\hat{q}_{IJ}(v)\cdot T^{v,s,{\rm norm}}_{\gamma,\vec{j},\vec{i}}(A)&=Q_{IJ}T^{v,s,{\rm norm}}_{\gamma,\vec{j},\vec{i}}(A)\,,\\
\widehat{V^{-1}}_{{\rm alt},v}\cdot T^{v,s,{\rm norm}}_{\gamma,\vec{j},\vec{i}}(A)&=\epsilon^{IJK}\epsilon^{LMN}Q_{IL}Q_{JM}Q_{KN}T^{v,s,{\rm norm}}_{\gamma,\vec{j},\vec{i}}(A)\,,\label{inverse-volume-eigen}
\end{align}
where
\begin{align}
Q_{IJ}&=\left(T^{v,s,{\rm norm}}_{\gamma,\vec{j},\vec{i}},\hat{q}_{IJ}(v)\cdot T^{v,s,{\rm norm}}_{\gamma,\vec{j},\vec{i}}\right)_{{\cal H}_{\rm kin}}=-\left({}^{(\frac12)}\!\hat{e}^\mu_I(v)\cdot T^{v,s,{\rm norm}}_{\gamma,\vec{j},\vec{i}},{}^{(\frac12)}\!\hat{e}^\mu_J(v)\cdot T^{v,s,{\rm norm}}_{\gamma,\vec{j},\vec{i}}\right)_{{\cal H}_{\rm kin}}\,.
\end{align}
In order to obtain the eigenvalues $Q_{IJ}$, we need to calculate the action of ${}^{(\frac12)}\!\hat{e}^\mu_I(v)$ on $T^{v,s,{\rm norm}}_{\gamma,\vec{j},\vec{i}}(A)$ or $T^{v,s}_{\gamma,\vec{j},\vec{i}}(A)$. In what follows, we only display the derivation of the two quantities $Q_{11}$ and $Q_{12}$, and the remaining components of $Q_{IJ}$ can be written down similarly.

Now let us consider the action of ${}^{(\frac12)}\!\hat{e}_1^\mu(v)$ on $T^{v,s}_{\gamma,\vec{j},\vec{i}}(A)$. Notice that the spherical tensors $\tau_\mu$ can be represented by (see Appendix A in \cite{graph-I})
\begin{align}\label{spher-rep-graph}
{[\pi_j(\tau_\mu)]^A}_B&=\frac{i}{2}\sqrt{2j(2j+1)(2j+2)}\makeSymbol{
\includegraphics[width=1.5cm]{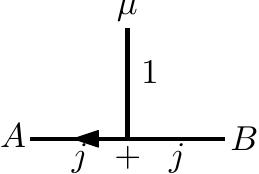}}=\frac{i}{2}\sqrt{2j(2j+1)(2j+2)}\makeSymbol{
\includegraphics[width=1.5cm]{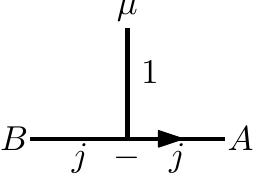}}\,.
\end{align}
Hence we have
\begin{align}\label{e-mu-1}
&{}^{(\frac12)}\!\hat{e}_1^\mu(v)\left[(-1)^{2j_3}\makeSymbol{
\includegraphics[width=3cm]{graph/Hamiltonian/graph-snf-1}}\right]={(\tau_\mu)^A}_B{[h_{s_1}]^B}_C\hat{V}^\frac12{[h_{s_1}^{-1}]^C}_A\left[(-1)^{2j_3}\makeSymbol{
\includegraphics[width=3cm]{graph/Hamiltonian/graph-snf-1}}\right]\notag\\
&={(\tau_\mu)^A}_B\sum_{j'_1}\left[V(j'_1,j_2,j_3)\right]^\frac12(-1)^{2j_3}d_{j'_1}\makeSymbol{
\includegraphics[width=3cm]{graph/Hamiltonian/graph-snf-7}}
=i\frac{\sqrt{6}}{2}\sum_{j'_1}\left[V(j'_1,j_2,j_3)\right]^\frac12(-1)^{2j_3}d_{j'_1}(-1)^{j_1+j'_1+\frac12}\makeSymbol{
\includegraphics[width=3cm]{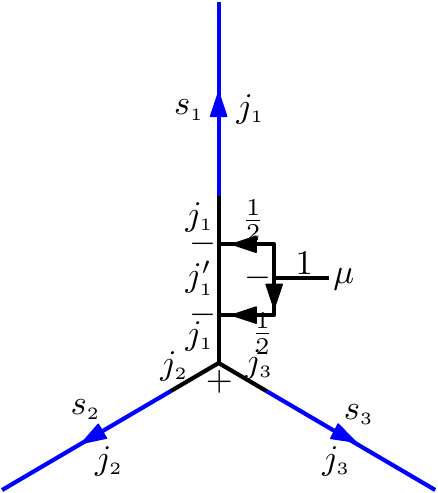}}\notag\\
&=i\frac{\sqrt{6}}{2}\sum_{j'_1}\left[V(j'_1,j_2,j_3)\right]^\frac12(-1)^{2j_3}d_{j'_1}(-1)^{j_1+j'_1-\frac12}
\begin{Bmatrix}
j'_1 & \frac12 & j_1\\
1 & j_1 & \frac12
\end{Bmatrix}\makeSymbol{
\includegraphics[width=3cm]{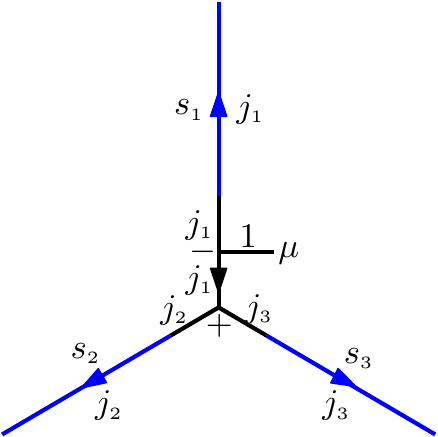}}\,,
\end{align}
where in the second step we have used the result of Eq. \eqref{snf-after-action}, and in the fourth step used the identity (see \ref{Appendix-identity-graph-proof} for proof)
\begin{align}\label{triad-6j-1}
\makeSymbol{
\includegraphics[width=1.2cm]{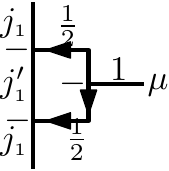}}&=-\begin{Bmatrix}
j'_1 & \frac12 & j_1\\
1 & j_1 & \frac12
\end{Bmatrix}\makeSymbol{
\includegraphics[width=1cm]{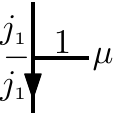}}\,.
\end{align}
Taking account of
\begin{align}
d_{j'_1}(-1)^{j_1+j'_1-\frac12}\begin{Bmatrix}
j'_1 & \frac12 & j_1\\
1 & j_1 & \frac12
\end{Bmatrix}&=-\frac{2}{\sqrt{6}}\sqrt{\frac{j_1(j_1+1)}{2j_1+1}}
\times\begin{cases}
1, & j'_1=j_1+\frac12\\
-1, & j'_1=j_1-\frac12
\end{cases}\,,\notag
\end{align}
Eq. \eqref{e-mu-1} can be reduced to
\begin{align}\label{e-mu-2}
&{}^{(\frac12)}\!\hat{e}_1^\mu(v)\left[(-1)^{2j_3}\makeSymbol{
\includegraphics[width=3cm]{graph/Hamiltonian/graph-snf-1}}\right]=-i\frac{\sqrt{j_1(j_1+1)}}{2j_1+1}\left(V^\frac12_{1A}-V^\frac12_{1B}\right)\left[\sqrt{d_{j_1}}(-1)^{2j_3}\makeSymbol{
\includegraphics[width=3cm]{graph/triad-like/triad-like-1}}\right]\,,
\end{align}
where
\begin{align}
V^\frac12_{1A}&:=\left[V(j'_1=j_1+1/2,j_2,j_3)\right]^\frac12,\qquad
V^\frac12_{1B}:=\left[V(j'_1=j_1-1/2,j_2,j_3)\right]^\frac12\,.
\end{align}
The intertwiner in Eq. \eqref{e-mu-2} is normalized because of
\begin{align}\label{intertwiner-e-1}
\sqrt{d_{j_1}}\,(-1)^{2j_3}\makeSymbol{
\includegraphics[width=1.2cm]{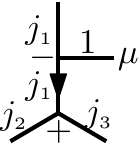}}=\sqrt{d_{j_1}d_{j_3}}\makeSymbol{
\includegraphics[width=4cm]{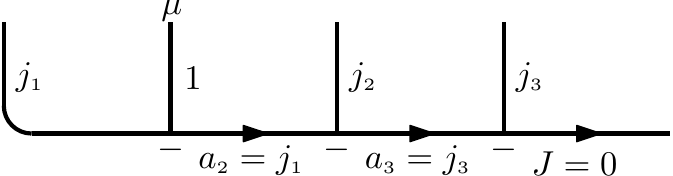}}\,.
\end{align}
Eq. \eqref{e-mu-2} implies that $\hat{e}_1^\mu(v)$ changes neither the graph nor the spins of $T^{v,s}_{\gamma,\vec{j},\vec{i}}(A)$ \eqref{graph-snf-v-s}. But it does change the intertwiner $i_v\equiv i^{\,J=0;\,a_2=j_3}_{j_1,j_2,j_3}$ into $i'_v\equiv i^{J=0;\,a_2=j_1,a_3=j_3}_{j_1,1,j_2,j_3}$ associated to $v$. Hence we obtain
\begin{align}\label{result-Q-11}
Q_{11}&=-\left({}^{(\frac12)}\!\hat{e}^\mu_1(v)T^{v,s,{\rm norm}}_{\gamma,\vec{j},\vec{i}},{}^{(\frac12)}\!\hat{e}^\mu_1(v)T^{v,s,{\rm norm}}_{\gamma,\vec{j},\vec{i}}\right)_{{\cal H}_{kin}}
=:-\left(-i\frac{\sqrt{j_1(j_1+1)}}{2j_1+1}\left(V^\frac12_{1A}-V^\frac12_{1B}\right)T^{v,s,{\rm norm}}_{\gamma,\vec{j},\vec{i}'},-i\frac{\sqrt{j_1(j_1+1)}}{2j_1+1}\left(V^\frac12_{1A}-V^\frac12_{1B}\right)T^{v,s,{\rm norm}}_{\gamma,\vec{j},\vec{i}'}\right)_{{\cal H}_{kin}}\notag\\
&=-\frac{j_1(j_1+1)}{(2j_1+1)^2}\left(V^\frac12_{1A}-V^\frac12_{1B}\right)^2 \int_{SU(2)^3}\prod_{I=1,2,3}{\rm d}\mu_H(h_{s_I})\overline{T^{v,s,{\rm norm}}_{\gamma,\vec{j},\vec{i}'}(A)}\,T^{v,s,{\rm norm}}_{\gamma,\vec{j},\vec{i}'}(A)\notag\\
&=-\frac{j_1(j_1+1)}{(2j_1+1)^2}\left(V^\frac12_{1A}-V^\frac12_{1B}\right)^2 {\rm tr}\left(\overline{i^{J=0;\,a_2=j_1,a_3=j_3}_{j_1,1,j_2,j_3}}\cdot i^{J=0;\,a_2=j_1,a_3=j_3}_{j_1,1,j_2,j_3}\right)\notag\\
&=-\frac{j_1(j_1+1)}{(2j_1+1)^2}\left(V^\frac12_{1A}-V^\frac12_{1B}\right)^2 {\rm tr}\left(i^{J=0;\,a_2=j_1,a_3=j_3}_{j_1,1,j_2,j_3}\cdot i^{J=0;\,a_2=j_1,a_3=j_3}_{j_1,1,j_2,j_3}\right)\notag\\
&=-\frac{j_1(j_1+1)}{(2j_1+1)^2}\left(V^\frac12_{1A}-V^\frac12_{1B}\right)^2\,,
\end{align}
where ${\rm tr}(\,)$ denotes contracting magnetic quantum numbers, we have integrated holonomies to give the contraction of the intertwiner with its complex conjugate in the fourth step, used the fact that the intertwiner is real in the fifth step, and the intertwiner is normalized in the last step.

Similarly, the action of $\hat{e}_2^\mu(v)$ on $T^{v,s}_{\gamma,\vec{j},\vec{i}}(A)$ yields
\begin{align}
{}^{(\frac12)}\!\hat{e}_2^\mu(v)\left[(-1)^{2j_3}\makeSymbol{
\includegraphics[width=3cm]{graph/Hamiltonian/graph-snf-1}}\right]&=-i\frac{\sqrt{j_2(j_2+1)}}{2j_2+1}\left(V^\frac12_{2A}-V^\frac12_{2B}\right)\left[\sqrt{2j_2+1}(-1)^{2j_3}\makeSymbol{
\includegraphics[width=3.2cm]{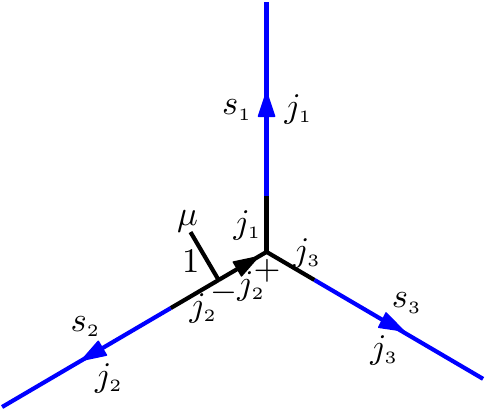}}\right]\notag\\
&=-i\frac{\sqrt{j_2(j_2+1)}}{2j_2+1}\left(V^\frac12_{2A}-V^\frac12_{2B}\right)(-1)^{j_1+j_2+j_3}\sum_a\sqrt{(2a+1)(2j_2+1)}
\begin{Bmatrix}
j_3 & j_2 & a\\
1 & j_1 & j_2
\end{Bmatrix}\notag\\
&\qquad\times\left[\sqrt{2a+1}(-1)^{2j_3}\makeSymbol{
\includegraphics[width=3cm]{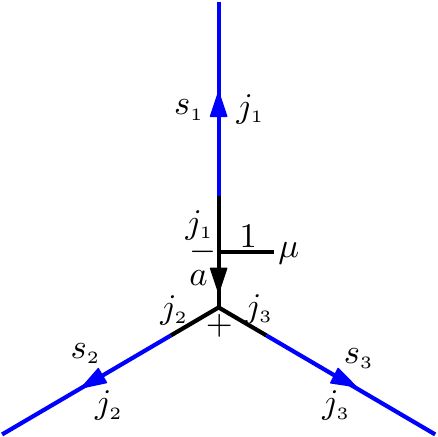}}\right]\,,
\end{align}
where
\begin{align}
V^\frac12_{2A}&:=\left[V(j'_2=j_2+1/2,j_1,j_3)\right]^\frac12,\qquad
V^\frac12_{2B}:=\left[V(j'_2=j_2-1/2,j_1,j_3)\right]^\frac12\,,
\end{align}
and in the second step we have used the following identity (see \ref{Appendix-identity-graph-proof} for proof)
\begin{align}\label{triad-6j-2}
\makeSymbol{
\includegraphics[width=1.5cm]{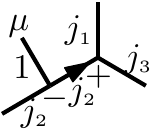}}&=(-1)^{j_1+j_2+j_3}\sum_a(2a+1)
\begin{Bmatrix}
j_3 & j_2 & a\\
1 & j_1 & j_2
\end{Bmatrix}\makeSymbol{
\includegraphics[width=1.2cm]{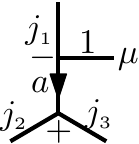}}\,.
\end{align}
Hence ${}^{(\frac12)}\!\hat{e}_2^\mu(v)$ changes $i_v\equiv i^{J=0;\,a_2=j_3}_{j_1,j_2,j_3}$ into composition of $i''_v\equiv i^{J=0;\,a_2=a,a_3=j_3}_{j_1,1,j_2,j_3}$ associated to $v$. Finally, we have
\begin{align}\label{result-Q-12}
Q_{12}&=-\left({}^{(\frac12)}\!\hat{e}^\mu_1(v)\cdot T^{v,s,{\rm norm}}_{\gamma,\vec{j},\vec{i}},{}^{(\frac12)}\!\hat{e}^\mu_2(v)\cdot T^{v,s,{\rm norm}}_{\gamma,\vec{j},\vec{i}}\right)_{{\cal H}_{kin}}\notag\\
&=-\frac{\sqrt{j_1(j_1+1)}}{2j_1+1}\left(V^\frac12_{1A}-V^\frac12_{1B}\right)\frac{\sqrt{j_2(j_2+1)}}{2j_2+1}\left(V^\frac12_{2A}-V^\frac12_{2B}\right)(-1)^{j_1+j_2+j_3}\notag\\
&\hspace{3cm}\times\sum_a\sqrt{(2a+1)(2j_2+1)}
\begin{Bmatrix}
j_3 & j_2 & a\\
1 & j_1 & j_2
\end{Bmatrix}{\rm tr}\left(i^{J=0;\,a_2=j_1,a_3=j_3}_{j_1,1,j_2,j_3}\cdot i^{J=0;\,a_2=a,a_3=j_3}_{j_1,1,j_2,j_3}\right)\notag\\
&=-\frac{\sqrt{j_1(j_1+1)j_2(j_2+1)}}{(2j_1+1)(2j_2+1)}\left(V^\frac12_{1A}-V^\frac12_{1B}\right)\left(V^\frac12_{2A}-V^\frac12_{2B}\right)(-1)^{j_1+j_2+j_3}\sum_a\sqrt{(2a+1)(2j_2+1)}
\begin{Bmatrix}
j_3 & j_2 & a\\
1 & j_1 & j_2
\end{Bmatrix}\delta_{a,j_1}\notag\\
&=-\frac{\sqrt{j_1(j_1+1)j_2(j_2+1)}}{(2j_1+1)(2j_2+1)}\left(V^\frac12_{1A}-V^\frac12_{1B}\right)\left(V^\frac12_{2A}-V^\frac12_{2B}\right)(-1)^{j_1+j_2+j_3}\sqrt{(2j_1+1)(2j_2+1)}
\begin{Bmatrix}
j_3 & j_2 & j_1\\
1 & j_1 & j_2
\end{Bmatrix}\notag\\
&=-\frac{\sqrt{j_1(j_1+1)j_2(j_2+1)}}{(2j_1+1)(2j_2+1)}\left(V^\frac12_{1A}-V^\frac12_{1B}\right)\left(V^\frac12_{2A}-V^\frac12_{2B}\right)(-1)^{j_1+j_2+j_3}\sqrt{(2j_1+1)(2j_2+1)}\notag\\
&\qquad\times(-1)^{j_1+j_2+j_3+1}
\frac{2[j_1(j_1+1)+j_2(j_2+1)-j_3(j_3+1)]}{\sqrt{2j_1(2j_1+1)(2j_1+2)2j_2(2j_2+1)(2j_2+2)}}\notag\\
&=\frac{j_1(j_1+1)+j_2(j_2+1)-j_3(j_3+1)}{2(2j_1+1)(2j_2+1)}\left(V^\frac12_{1A}-V^\frac12_{1B}\right)\left(V^\frac12_{2A}-V^\frac12_{2B}\right)\,.
\end{align}
Similarly we can write down the remaining components of $Q_{IJ}$ and thus the eigenvalue of $\widehat{V^{-1}}_{{\rm alt},v}$ in \eqref{inverse-volume-eigen}.

\section{Summary}\label{sec-IV}
In this paper, the graphical method is employed to compute explicitly the actions of the Euclidean Hamiltonian constraint and inverse volume operators on the spin network states with trivalent vertices. The rules of transforming graphs in our method simplify greatly the calculation. Every graphical reduction in our graphical calculus corresponds to an algebraic reduction in an unique and unambiguous way. This ensures the rigour of our calculation.

The action of the Euclidean Hamiltonian $\hat{H}^E_v$ on the spin network states $T^{v,s}_{\gamma,\vec{j},\vec{i}}(A)$ with trivalent vertex $v$ was shown in \eqref{action-H-E-v}. The difference between our result \eqref{H-IJK}) and (II.14) in \cite{Alesci:2011ia} is the factor $-1/2$. Based on the gauge invariant operators $\hat{q}_{IJ}(v)$, the inverse volume operator $\widehat{V^{-1}}_{{\rm alt},v}$ defined in \eqref{inverse-volume-operator} takes eigenvalues on the orthonormal spin network state $T^{v,s,{\rm norm}}_{\gamma,\vec{j},\vec{i}}(A)$. The eigenvalues of $\widehat{V^{-1}}_{{\rm alt},v}$, presented in \eqref{inverse-volume-eigen}, consist of the eigenvalues $Q_{IJ}$ of $\hat{q}_{IJ}(v)$. The eigenvalues of $\widehat{V^{-1}}_{{\rm alt},v}$ were also derived in \cite{Brunnemann:2005ip} by the algebraic method. The difference between the value of $Q_{11}$ in \eqref{result-Q-11} and (4.17) in \cite{Brunnemann:2005ip} is a minus sign, while the difference between the value of $Q_{12}$ in \eqref{result-Q-12} and (4.17) in \cite{Brunnemann:2005ip} is the factor $-[2+(-1)^{2(j_1+j_2)}]/3$.

In principle, the graphical calculation method can be applied to the general cases where the spin network states are defined on arbitrary valent vertices and the holonomies appeared in the two operators are expressed in arbitrary representation of the gauge group. However, for those general cases, the volume operator lacks the explicit matrix elements formula. This prevents us from doing further calculation. For the same reason, the matrix elements of the Lorentzian part contained in the full gravitational Hamiltonian constraint operator have not been explicitly written down even on the trivalent vertices \cite{Alesci:2013kpa}.

\section*{Acknowledgments}
The authors would like to thank Antonia Zipfel for helpful discussions. J. Y. would also like to thank Chopin Soo and Hoi-Lai Yu for useful discussions. J. Y. is supported in part by NSFC No. 11347006, by the Institute of Physics, Academia Sinica, Taiwan, and by the Natural Science Foundation of Guizhou University (No. 47 in 2013). Y. M. is supported in part by the NSFC (Grant Nos. 11235003 and 11475023) and the Research Fund for the Doctoral Program of Higher Education of China.

\appendix
\renewcommand\thesection{\appendixname~\Alph{section}}
\renewcommand\thesubsection{\Alph{section}.\arabic{subsection}}
\renewcommand\theequation{\Alph{section}.\arabic{equation}}

\section{The diagonalization of volume operator in 2-d intertwiner space}\label{volume-eigenvalue}
We denote
\begin{align}
|\alpha_1\rangle\equiv|1/2,j'_1,j_2,j_3;a_2=j'_1-1/2,a_3=j_3,J=0\rangle\equiv\sqrt{2a_2+1}\sqrt{2a_3+1}\makeSymbol{
\includegraphics[width=4cm]{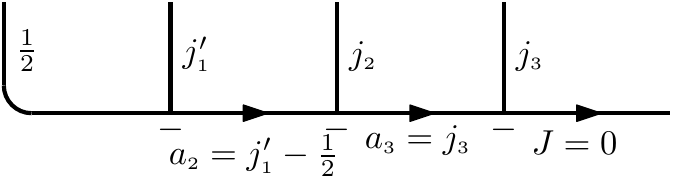}}\,,\\
|\alpha_2\rangle\equiv|1/2,j'_1,j_2,j_3;a_2=j'_1+1/2,a_3=j_3,J=0\rangle\equiv\sqrt{2a_2+1}\sqrt{2a_3+1}\makeSymbol{
\includegraphics[width=4cm]{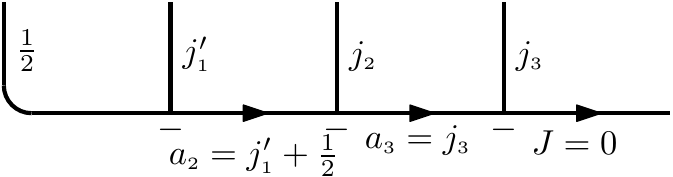}}\,.
\end{align}
The matrix of the operator $i\hat{q}_{j'_1j_2j_3}$ in the above two states reads
\begin{align}\label{q-1-2-3-matrix}
(i\hat{q}_{j'_1j_2j_3})=
\begin{pmatrix}
\langle\alpha_1|i\hat{q}_{j'_1j_2j_3}|\alpha_1\rangle & \langle\alpha_1|i\hat{q}_{j'_1j_2j_3}|\alpha_2\rangle\\
\langle\alpha_2|i\hat{q}_{j'_1j_2j_3}|\alpha_1\rangle & \langle\alpha_2|i\hat{q}_{j'_1j_2j_3}|\alpha_2\rangle
\end{pmatrix}=\begin{pmatrix}
0 & \langle\alpha_1|i\hat{q}_{j'_1j_2j_3}|\alpha_2\rangle\\
\langle\alpha_2|i\hat{q}_{j'_1j_2j_3}|\alpha_1\rangle & 0
\end{pmatrix}=:\begin{pmatrix}
0 & -ib\\
ib & 0
\end{pmatrix}\,.
\end{align}
The eigenvalues and corresponding (normalized) eigenvectors of $i\hat{q}_{j'_1j_2j_3}$ are given by
\begin{align}
\lambda_1=-b\rightarrow e_1=\frac{1}{\sqrt{2}}
\begin{pmatrix}
i\\
1
\end{pmatrix},\qquad
\lambda_2=b\rightarrow e_2=\frac{1}{\sqrt{2}}
\begin{pmatrix}
-i\\
1
\end{pmatrix}\,.
\end{align}
Hence we obtain
\begin{align}
\sqrt{|i\hat{q}_{j'_1j_2j_3}|}\;|\alpha_1\rangle&=\sum_{i=1}^2|e_i\rangle\langle e_i|\sqrt{|i\hat{q}_{j'_1j_2j_3}|}\;|\alpha_1\rangle=\sum_{i=1}^2\sqrt{|\lambda_i|}\,|e_i\rangle\langle e_i|\alpha_1\rangle
=\sqrt{|b|}\,|\alpha_1\rangle\,,\\
\sqrt{|i\hat{q}_{j'_1j_2j_3}|}\;|\alpha_2\rangle&=\sum_{i=1}^2|e_i\rangle\langle e_i|\sqrt{|i\hat{q}_{j'_1j_2j_3}|}\;|\alpha_2\rangle=\sum_{i=1}^2\sqrt{|\lambda_i|}\,|e_i\rangle\langle e_i|\alpha_2\rangle=\sqrt{|b|}\,|\alpha_2\rangle\,.
\end{align}
Now we derive the value of $|b|$. Using the matrix elements of $\langle \vec{a}'|\hat{q}_{234}|\vec{a}\rangle$ in \cite{graph-I}, we have
\begin{align}\label{q-123-a2}
\langle\alpha_1|\hat{q}_{j'_1j_2j_3}|\alpha_2\rangle\equiv&\langle a_2-1=j'_1-\frac12|\hat{q}_{j'_1j_2j_3}|a_2=j'_1+\frac12\rangle\notag\\
&=-\frac14(-1)^{\frac12+j'_1+j_3}(-1)^{j'_1+\frac12-(j'_1-\frac12)}X(j'_1,j_2)^{\frac12}X(j_2,j_3)^{\frac12}\sqrt{2j'_1(2j'_1+2)}(2j_3+1)
\begin{Bmatrix} \frac12 & j'_1 &  j'_1+\frac12 \\
 1 & j'_1-\frac12 & j'_1
\end{Bmatrix}\;\begin{Bmatrix}  0 & j_3 & j_3 \\
  1 & j_3 & j_3
\end{Bmatrix}\notag\\
&\quad\times\left[(-1)^{j'_1-\frac12+j_3}\begin{Bmatrix}  j_3 & j_2 & j'_1+\frac12 \\
  1 & j'_1-\frac12 & j_2
\end{Bmatrix}\begin{Bmatrix} j'_1-\frac12 & j_2 &  j_3 \\
 1 & j_3 & j_2
\end{Bmatrix}-(-1)^{j'_1+\frac12+j_3}\begin{Bmatrix}  j_3 & j_2 & j'_1+\frac12 \\
  1 & j'_1-\frac12 & j_2
\end{Bmatrix}\begin{Bmatrix} j'_1+\frac12 & j_2 &  j_3 \\
 1 & j_3 & j_2
\end{Bmatrix}
\right]\notag\\
&=-\frac14(-1)^{\frac12+j'_1+j_3}(-1)X(j'_1,j_2)^{\frac12}X(j_2,j_3)^{\frac12}\notag\\
&\quad\times\sqrt{2j'_1(2j'_1+2)}(2j_3+1)
\begin{Bmatrix} \frac12 & j'_1 &  j'_1+\frac12 \\
 1 & j'_1-\frac12 & j'_1
\end{Bmatrix}\;\begin{Bmatrix}  0 & j_3 & j_3 \\
  1 & j_3 & j_3
\end{Bmatrix}\;\begin{Bmatrix}  j_3 & j_2 & j'_1+\frac12 \\
  1 & j'_1-\frac12 & j_2
\end{Bmatrix}\notag\\
&\quad\times\left[(-1)^{j'_1-\frac12+j_3}\begin{Bmatrix} j'_1-\frac12 & j_2 &  j_3 \\
 1 & j_3 & j_2
\end{Bmatrix}-(-1)^{j'_1+\frac12+j_3}\begin{Bmatrix} j'_1+\frac12 & j_2 &  j_3 \\
 1 & j_3 & j_2
\end{Bmatrix}
\right]\,.
\end{align}
With the following values of $6j$-symbols
\begin{align}
\begin{Bmatrix} \frac12 & j'_1 &  j'_1+\frac12 \\
 1 & j'_1-\frac12 & j'_1
\end{Bmatrix}&=(-1)^{2j'_1+1}\left[\frac{2}{2j'_1(2j'_1+1)^2(2j'_1+2)}\right]^{\frac12}\,,\notag\\
\begin{Bmatrix}  0 & j_3 & j_3 \\
  1 & j_3 & j_3
\end{Bmatrix}&=\frac{(-1)^{2j_3+1}}{2j_3+1}\,,\notag\\
\begin{Bmatrix}  j_3 & j_2 & j'_1+\frac12 \\
  1 & j'_1-\frac12 & j_2
\end{Bmatrix}&=(-1)^{j'_1+j_2+j_3+\frac12}\left[\frac{2(j'_1+j_2+j_3+\frac32)(j'_1+j_2-j_3+\frac12)(j'_1-j_2+j_3+\frac12)(-j'_1+j_2+j_3+\frac12)}{X(j_2,j'_1)}\right]^{\frac12}\,,\notag\\
\begin{Bmatrix} j'_1-\frac12 & j_2 &  j_3 \\
 1 & j_3 & j_2
\end{Bmatrix}&=(-1)^{j'_1+j_2+j_3+\frac12}\frac{2[j_2(j_2+1)+j_3(j_3+1)-(j'_1-\frac12)(j'_1+\frac12)]}{X(j_2,j_3)^{1/2}}\,,\notag\\
\begin{Bmatrix} j'_1+\frac12 & j_2 &  j_3 \\
 1 & j_3 & j_2
\end{Bmatrix}&=(-1)^{j'_1+j_2+j_3+\frac32}\frac{2[j_2(j_2+1)+j_3(j_3+1)-(j'_1+\frac12)(j'_1+\frac32)]}{X(j_2,j_3)^{1/2}}\,,
\end{align}
where $X(j_1,j_2)\equiv 2j_1(2j_1+1)(2j_1+2)2j_2(2j_2+1)(2j_2+2)$, the matrix element formula \eqref{q-123-a2} can be simplified as
\begin{align}
&\langle a_2-1=j'_1-\frac12|\hat{q}_{j'_1j_2j_3}|a_2=j'_1+\frac12\rangle&=\left[(j'_1+j_2+j_3+\frac32)(j'_1+j_2-j_3+\frac12)(j'_1-j_2+j_3+\frac12)(-j'_1+j_2+j_3+\frac12)\right]^{\frac12}\,,
\end{align}
which implies that the absolute value of $b$ in the matrix elements \eqref{q-1-2-3-matrix} takes the form
\begin{align}
|b|=\left[(j'_1+j_2+j_3+\frac32)(j'_1+j_2-j_3+\frac12)(j'_1-j_2+j_3+\frac12)(-j'_1+j_2+j_3+\frac12)\right]^{\frac12}\,.
\end{align}
Therefore, we have
\begin{align}
\hat{V}\,|\alpha_i\rangle&=\frac{\ell_{\rm p}^3\,\beta^{3/2}}{4\sqrt{2}}\sqrt{|i\hat{q}_{j'_1j_2j_3}|}\,|\alpha_i\rangle=\frac{\ell_{\rm p}^3\,\beta^{3/2}}{4\sqrt{2}}\sqrt{|b|}\,|\alpha_i\rangle\notag\\
&=\frac{\ell_{\rm p}^3\,\beta^{3/2}}{4\sqrt{2}}\left[(j'_1+j_2+j_3+\frac32)(j'_1+j_2-j_3+\frac12)(j'_1-j_2+j_3+\frac12)(-j'_1+j_2+j_3+\frac12)\right]^{\frac14}\,|\alpha_i\rangle\notag\\
&\equiv V(1/2,j'_1,j_2,j_3;a_2=j'_1+1/2,a_3=j_3)\,|\alpha_i\rangle,\qquad i=1,2\,,
\end{align}
which reveals that the volume operator is diagonal in the 2-dimensional intertwiner space.

\section{Proof of some graphical identities}\label{Appendix-identity-graph-proof}
In the graphical calculus, one usually uses the following identity ((A.60) in \cite{graph-I}) to simplify graphs,
\begin{align}\label{6j-3j-relation-org}
\makeSymbol{
\includegraphics[width=2.6cm]{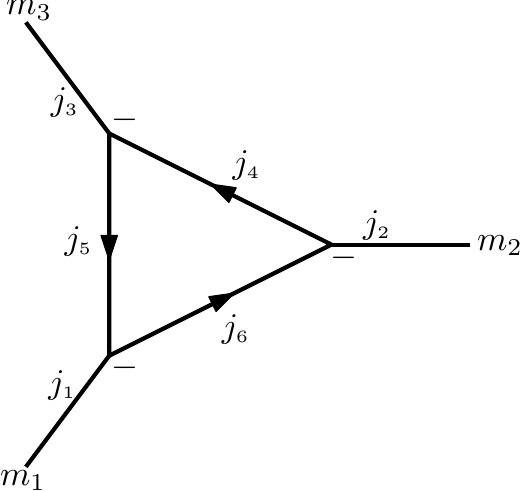}}=\makeSymbol{
\includegraphics[width=1.5cm]{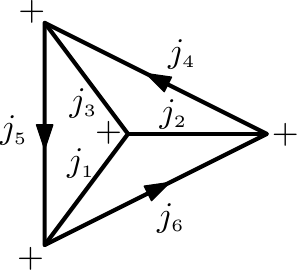}}\times\makeSymbol{
\includegraphics[width=1.5cm]{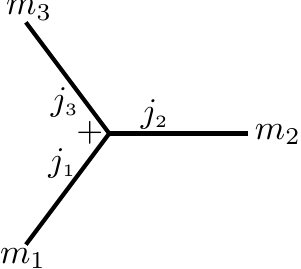}}\,.
\end{align}
The graph on the left-hand side of \eqref{6j-3j-relation-org} can be transformed to
\begin{align}
\makeSymbol{
\includegraphics[width=2.6cm]{graph/Hamiltonian/6j-3j-symbol}}=\makeSymbol{
\includegraphics[width=2.6cm]{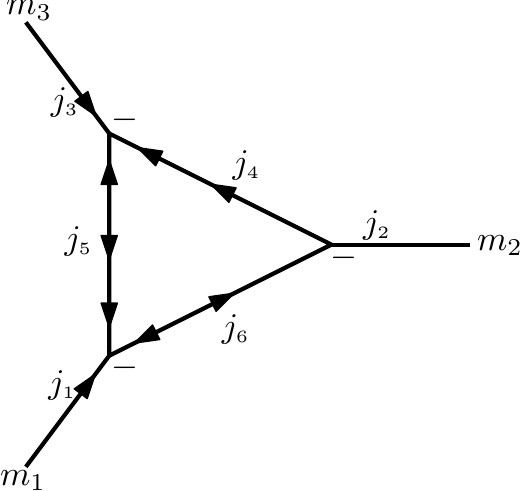}}=(-1)^{2j_4}\makeSymbol{
\includegraphics[width=2.6cm]{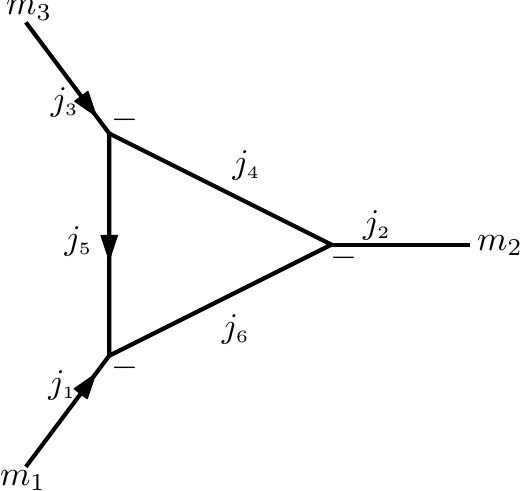}}=(-1)^{2(j_1+j_4)}\makeSymbol{
\includegraphics[width=2.6cm]{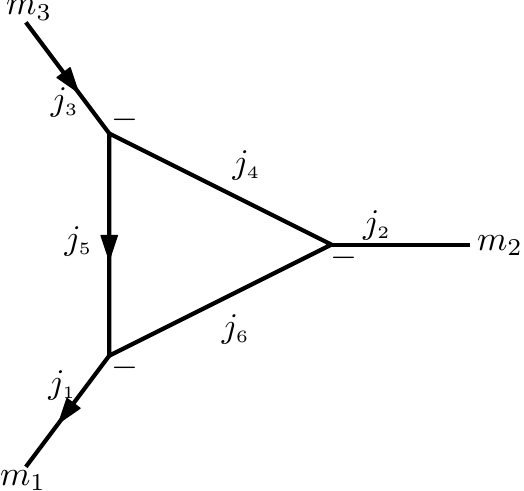}}\,.
\end{align}
The first graph on the right-hand side of \eqref{6j-3j-relation-org} represents the $6j$-symbol
\begin{align}
\makeSymbol{
\includegraphics[width=1.5cm]{graph/Hamiltonian/6j-symbol-def-2}}=
\begin{Bmatrix}
j_1 & j_2 & j_3\\
j_4 & j_5 & j_6
\end{Bmatrix}\,,
\end{align}
which can be transformed as
\begin{align}
\makeSymbol{
\includegraphics[width=1.5cm]{graph/Hamiltonian/6j-symbol-def-2}}&=\makeSymbol{
\includegraphics[width=1.5cm]{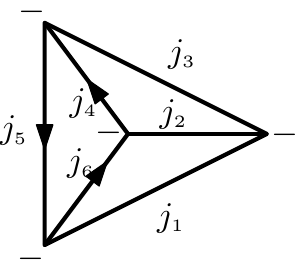}}=\makeSymbol{
\includegraphics[width=1.5cm]{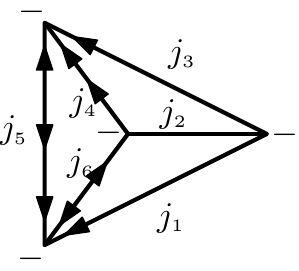}}=(-1)^{2j_4}\makeSymbol{
\includegraphics[width=1.5cm]{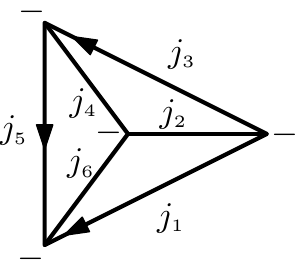}}=(-1)^{2(j_1+j_4)}\makeSymbol{
\includegraphics[width=1.5cm]{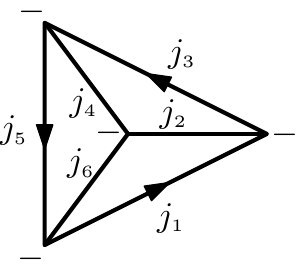}}\,,
\end{align}
where we have used the rules of transforming graphs ((A.50)-(A.53) in \cite{graph-I})
\begin{align}
\makeSymbol{
\includegraphics[width=2cm]{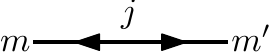}}&=
\makeSymbol{
\includegraphics[width=2cm]{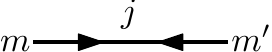}}=\makeSymbol{
\includegraphics[width=2cm]{graph/Hamiltonian/wigner-3j-symbol-9}}\,,\label{two-arrow-cancel}\\
\makeSymbol{
\includegraphics[width=2cm]{graph/Hamiltonian/wigner-3j-symbol-7}}&=\makeSymbol{
\includegraphics[width=2cm]{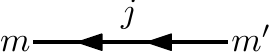}}=(-1)^{2j}\makeSymbol{
\includegraphics[width=2cm]{graph/Hamiltonian/wigner-3j-symbol-9}}\,,\label{two-arrow-result}\\
\makeSymbol{
\includegraphics[width=2cm]{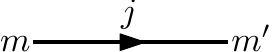}}&=(-1)^{2j}\makeSymbol{
\includegraphics[width=2cm]{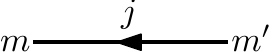}}\,,\label{arrow-flip}\\
\makeSymbol{
\includegraphics[width=2cm]{graph/Hamiltonian/wigner-3j-symbol-1}}&=\makeSymbol{
\includegraphics[width=2cm]{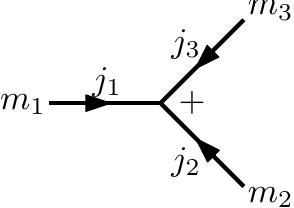}}=\makeSymbol{
\includegraphics[width=2cm]{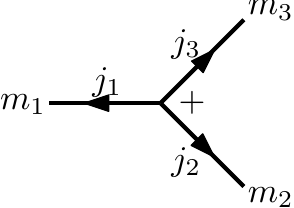}}\,.\label{three-arrow-adding}
\end{align}
Hence Eq. \eqref{6j-3j-relation-org} is equal to the following graphical identity
\begin{align}
\makeSymbol{
\includegraphics[width=2.6cm]{graph/Hamiltonian/6j-3j-symbol-id-3}}&=\makeSymbol{
\includegraphics[width=1.5cm]{graph/Hamiltonian/6j-3j-symbol-id-7}}\times\makeSymbol{
\includegraphics[width=1.5cm]{graph/Hamiltonian/6j-3j-symbol-1}}\,,
\end{align}
which implies
\begin{align}\label{6j-3j-relation-1}
\makeSymbol{
\includegraphics[width=2.6cm]{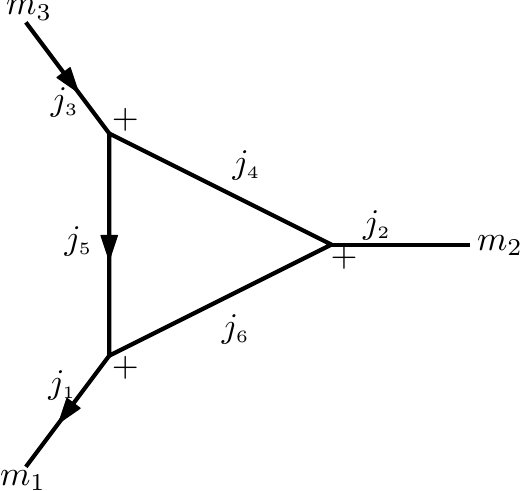}}&=\makeSymbol{
\includegraphics[width=1.5cm]{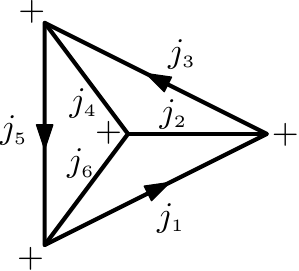}}\times\makeSymbol{
\includegraphics[width=1.5cm]{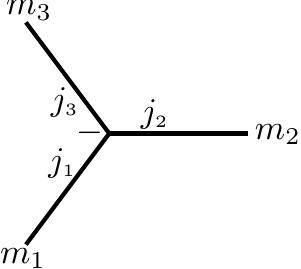}}
=(-1)^{j_1+j_2+j_3}\makeSymbol{
\includegraphics[width=1.5cm]{graph/Hamiltonian/6j-3j-symbol-id-9}}\times\makeSymbol{
\includegraphics[width=1.5cm]{graph/Hamiltonian/6j-3j-symbol-1}}\,.
\end{align}

Graphically, Eq. \eqref{ham-6j-1} can be proved by
\begin{align}
\makeSymbol{
\includegraphics[width=1.5cm]{graph/Hamiltonian/graph-origin-snf-loop-id-1-1}}&=\makeSymbol{
\includegraphics[width=1.5cm]{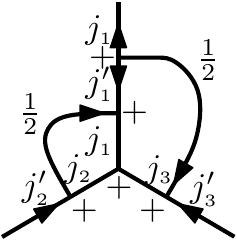}}=(-1)^{2\times\frac12+2j'_2}\makeSymbol{
\includegraphics[width=1.5cm]{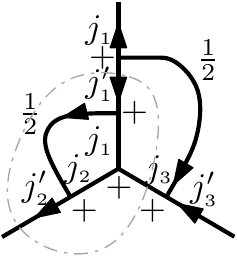}}=(-1)^{2\times\frac12+2j'_2}(-1)^{j'_1+j'_2+j_3}\makeSymbol{
\includegraphics[width=1.5cm]{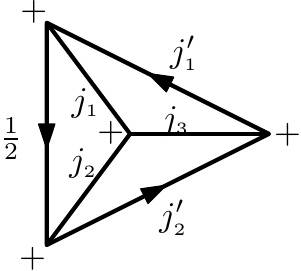}}\makeSymbol{
\includegraphics[width=1.5cm]{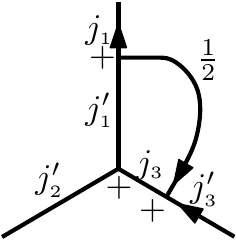}}\notag\\
&=(-1)^{\frac12\times2+2j'_2}(-1)^{j'_1+j'_2+j_3}(-1)^{2\times\frac12}\makeSymbol{
\includegraphics[width=1.5cm]{graph/Hamiltonian/graph-origin-snf-loop-id-1-4}}\makeSymbol{
\includegraphics[width=1.5cm]{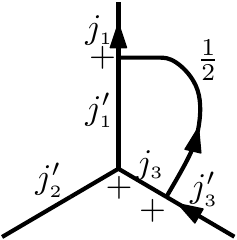}}\notag\\
&=(-1)^{\frac12\times2+2j'_2}(-1)^{j'_1+j'_2+j_3}(-1)^{2\times\frac12}(-1)^{j_1+j'_2+j'_3}\makeSymbol{
\includegraphics[width=1.5cm]{graph/Hamiltonian/graph-origin-snf-loop-id-1-4}}\makeSymbol{
\includegraphics[width=1.5cm]{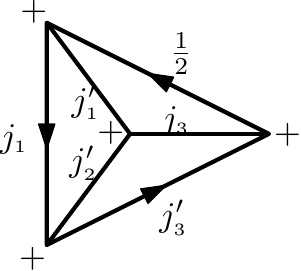}}\makeSymbol{
\includegraphics[width=1cm]{graph/Hamiltonian/graph-origin-snf-loop-id-1-8}}\notag\\
&=-(-1)^{j_1+j'_1+\frac12}(-1)^{j_3+j'_3+\frac12}
\begin{Bmatrix}
j_2 & j_3 & j_1\\
j'_1 & \frac12 & j'_2
\end{Bmatrix}
\begin{Bmatrix}
j'_2 & j_3 & j'_1\\
\frac12 & j_1 & j'_3
\end{Bmatrix}\makeSymbol{
\includegraphics[width=1cm]{graph/Hamiltonian/graph-origin-snf-loop-id-1-8}}\notag\\
&=-(-1)^{j_1+j'_1+\frac12}(-1)^{j_3+j'_3+\frac12}
\begin{Bmatrix}
j_1 & \frac12 & j'_1\\
j'_2 & j_3 & j_2
\end{Bmatrix}
\begin{Bmatrix}
j_1 & \frac12 & j'_1\\
j_3 & j'_2 & j'_3
\end{Bmatrix}\makeSymbol{
\includegraphics[width=1cm]{graph/Hamiltonian/graph-origin-snf-loop-id-1-8}}\,,
\end{align}
where we have used the rules \eqref{two-arrow-cancel}-\eqref{three-arrow-adding}, and \eqref{6j-3j-relation-1} in the third and fifth steps respectively. Similarly, Eq. \eqref{ham-6j-1} can be shown by
\begin{align}
\makeSymbol{
\includegraphics[width=1.5cm]{graph/Hamiltonian/graph-origin-snf-loop-id-2-1}}&=\makeSymbol{
\includegraphics[width=1.5cm]{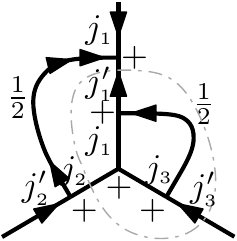}}=(-1)^{j'_1+j_2+j'_3}\makeSymbol{
\includegraphics[width=1.5cm]{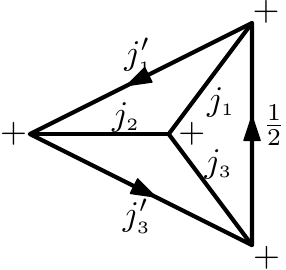}}\makeSymbol{
\includegraphics[width=1.5cm]{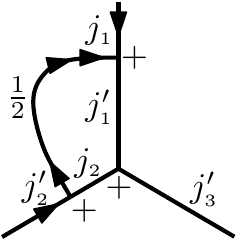}}
=(-1)^{j'_1+j_2+j'_3}(-1)^{2j'_2}\makeSymbol{
\includegraphics[width=1.5cm]{graph/Hamiltonian/graph-origin-snf-loop-id-2-3}}\makeSymbol{
\includegraphics[width=1.5cm]{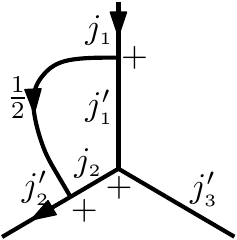}}\notag\\
&=(-1)^{j'_1+j_2+j'_3}(-1)^{2j'_2}(-1)^{j_1+j'_2+j'_3}\makeSymbol{
\includegraphics[width=1.5cm]{graph/Hamiltonian/graph-origin-snf-loop-id-2-3}}\makeSymbol{
\includegraphics[width=1.5cm]{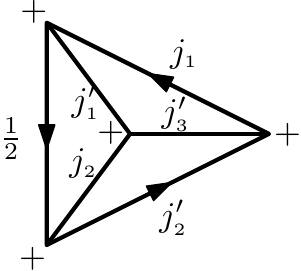}}\makeSymbol{
\includegraphics[width=1cm]{graph/Hamiltonian/graph-origin-snf-loop-id-1-8}}\notag\\
&=(-1)^{j'_1+j_2+j'_3}(-1)^{2j'_2}(-1)^{j_1+j'_2+j'_3}(-1)^{-2j'_2-2j'_3-2j_1}\begin{Bmatrix}
j_3 & j_1 & j_2\\
j'_1 & j'_3 & \frac12
\end{Bmatrix}\begin{Bmatrix}
j_2 & j'_3 & j'_1\\
j_1 & \frac12 & j'_2
\end{Bmatrix}\makeSymbol{
\includegraphics[width=1cm]{graph/Hamiltonian/graph-origin-snf-loop-id-1-8}}\notag\\
&=(-1)^{j_1-j'_1+\frac12}(-1)^{j_2+j'_2+\frac12}\begin{Bmatrix}
j_1 & \frac12 & j'_1\\
j_2 & j'_3 & j'_2
\end{Bmatrix}
\begin{Bmatrix}
j_1 & \frac12 & j'_1\\
j'_3 & j_2 & j_3
\end{Bmatrix}\makeSymbol{
\includegraphics[width=1cm]{graph/Hamiltonian/graph-origin-snf-loop-id-1-8}}\,,
\end{align}
where we have used \eqref{6j-3j-relation-1} in the second and fourth steps, and used the fact that the allowed triple $(j'_2,j'_3,j_1)$ satisfy the triangular condition in the fifth step.

The identity \eqref{triad-6j-1} can be proved by
\begin{align}
\makeSymbol{
\includegraphics[width=1.2cm]{graph/triad-like/triad-like-identity-1}}&=\makeSymbol{
\includegraphics[width=1.2cm]{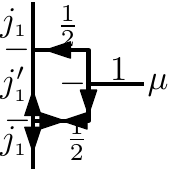}}=\makeSymbol{
\includegraphics[width=1.2cm]{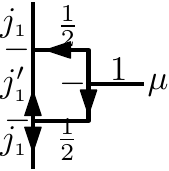}}=(-1)^{2j'_1}(-1)^{2\times\frac12}\makeSymbol{
\includegraphics[width=1.2cm]{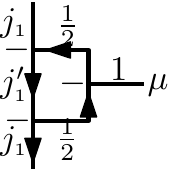}}=(-1)^{2j'_1+1}\makeSymbol{
\includegraphics[width=1.4cm]{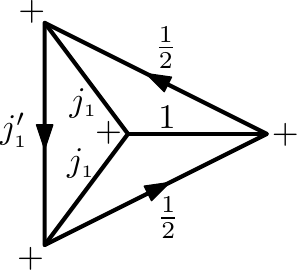}}\times\makeSymbol{
\includegraphics[width=1.2cm]{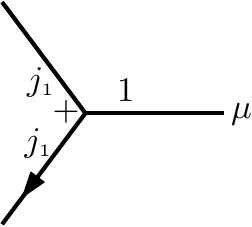}}\notag\\
&=(-1)^{2j'_1+1}
\begin{Bmatrix}
\frac12 & 1 & \frac12\\
j_1 & j'_1 & j_1
\end{Bmatrix}\makeSymbol{
\includegraphics[width=1cm]{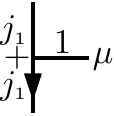}}=(-1)^{2j'_1+1}
\begin{Bmatrix}
j'_1 & \frac12 & j_1\\
1 & j_1 & \frac12
\end{Bmatrix}\makeSymbol{
\includegraphics[width=1cm]{graph/triad-like/triad-like-identity-7}}\notag\\
&=(-1)^{2j'_1+1}(-1)^{2j_1+1}
\begin{Bmatrix}
j'_1 & \frac12 & j_1\\
1 & j_1 & \frac12
\end{Bmatrix}\makeSymbol{
\includegraphics[width=1cm]{graph/triad-like/triad-like-identity-8}}
=(-1)^{2j'_1+2j_1+2}
\begin{Bmatrix}
j'_1 & \frac12 & j_1\\
1 & j_1 & \frac12
\end{Bmatrix}\makeSymbol{
\includegraphics[width=1cm]{graph/triad-like/triad-like-identity-8}}\notag\\
&=-\begin{Bmatrix}
j'_1 & \frac12 & j_1\\
1 & j_1 & \frac12
\end{Bmatrix}\makeSymbol{
\includegraphics[width=1cm]{graph/triad-like/triad-like-identity-8}}\,,
\end{align}
where we have used \eqref{6j-3j-relation-org} in fourth step, and used the fact that $(-1)^{2j'_1+2j_1+1}=1$ in the last step, since the allowed triple $(j'_1,j_1,\frac12)$ satisfy the triangular condition.

Eq. \eqref{triad-6j-2} can be obtained from
\begin{align}
\makeSymbol{
\includegraphics[width=1.2cm]{graph/triad-like/triad-like-intertwiner-3}}&=\makeSymbol{
\includegraphics[width=1.4cm]{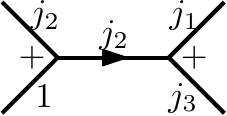}}=\sum_{a}(2a+1)(-1)^{1+j_3+j_2+a}
\begin{Bmatrix}
j_2 & j_3 & a\\
j_1 & 1 & j_2
\end{Bmatrix}
\makeSymbol{
\includegraphics[width=1.4cm]{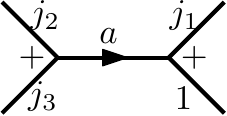}}\notag\\
&=\sum_{a}(2a+1)(-1)^{1+j_3+j_2+a}
\begin{Bmatrix}
j_2 & j_3 & a\\
j_1 & 1 & j_2
\end{Bmatrix}(-1)^{j_1+a+1}(-1)^{2a}
\makeSymbol{
\includegraphics[width=1.4cm]{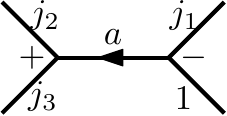}}\notag\\
&=(-1)^{j_1+j_2+j_3}\sum_a(2a+1)
\begin{Bmatrix}
j_3 & j_2 & a\\
1 & j_1 & j_2
\end{Bmatrix}\makeSymbol{
\includegraphics[width=1.2cm]{graph/triad-like/triad-like-intertwiner-7}}\,,
\end{align}
where in the second step we have used the identity ((A.62) in \cite{graph-I})
\begin{align}\label{6j-interchange}
\makeSymbol{
\includegraphics[width=2cm]{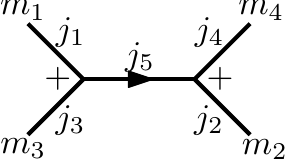}}&=\sum_{j_6}(2j_6+1)(-1)^{j_2+j_3+j_5+j_6}
\begin{Bmatrix}
j_1 & j_2 & j_6\\
j_4 & j_3 & j_5
\end{Bmatrix}
\makeSymbol{
\includegraphics[width=2cm]{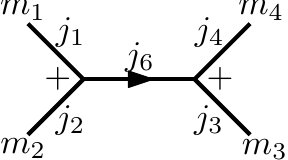}}\,.
\end{align}

\providecommand{\href}[2]{#2}\begingroup\raggedright\endgroup


\end{document}